\documentclass{article}
\usepackage{amsmath,amsfonts,setspace,cancel,url,hyperref,verbatim,soul,cite}
\usepackage[a4paper]{geometry}
\newcommand{\be}{\begin{equation}}
\newcommand{\ee}{\end{equation}}

\newcommand{\gM}{\mathcal{H}}
\newcommand{\tz}{\tilde z}
\newcommand{\Cc}{\mathcal{C}}
\newcommand{\Dop}{\slash\!\!\!\!\mathcal{D}}
\numberwithin{equation}{section}
\begin{document}

\begin{titlepage}

\begin{center}

\hfill DAMTP-2015-42

\vskip 1.5cm

{\LARGE \sc  Conserved currents of double field theory}

\vskip 1cm

{\large \sc Chris D. A. Blair} \\

\vskip 25pt

{\em  Department of Applied Mathematics and Theoretical Physics,\\Centre for Mathematical Sciences, University of Cambridge, \\
Wilberforce Road, Cambridge CB3 0WA, United Kingdom.
\\
 \vspace{1em}
\vskip 5pt }

{{\tt C.D.A.Blair@damtp.cam.ac.uk}}\,{}
\\

\end{center}

\vskip 0.5cm

\begin{center} {\bf Abstract}\\[3ex]
\end{center}

\noindent 
We find the conserved current associated to invariance under generalised diffeomorphisms in double field theory. This can be used to define a generalised Komar integral. We comment on its applications to solutions, in particular to the fundamental string/pp-wave. We also discuss the current in the context of Scherk-Schwarz compactifications. We calculate the current for both the original double field theory action, corresponding to the NSNS sector alone, and for the RR sector.

\thispagestyle{empty}
\setcounter{footnote}{0}

\end{titlepage}


\section{Introduction}

\subsection{General introduction}

A central focus of physics is the study of symmetry and its consequences. The symmetries
of the low energy limit of string theory - supergravity - include diffeomorphisms, gauge
transformations, supersymmetry and, due to their stringy origin, duality symmetries. 

An important example of the latter is T-duality. 
At the worldsheet level, it
arises from the ability to exchange the momentum and winding of a wrapped string. 
In the NSNS sector of supergravity, this relates
different backgrounds involving the metric, $B$-field and dilaton. 

This paper forms part of a line of research into reformulations of string theory and supergravity
which treat T-duality as a manifest symmetry. This can be achieved either by using generalised
geometry 
\cite{Hitchin:2004ut, Gualtieri:2003dx,
Coimbra:2011nw, Coimbra:2012yy
}, in which T-duality is realised as a symmetry of an extended
tangent bundle, or by introducing extra coordinates, thought of as dual to winding, so that one
studies a novel doubled geometry. This builds on the pioneering work of 
\cite{Duff:1989tf, Tseytlin:1990nb, Tseytlin:1990va, Siegel:1993xq,Siegel:1993th}. More recent
efforts have led to the introduction of ``double field theory'' which in its original formulation
defines a T-duality covariant description of the NSNS sector on a doubled space
\cite{
Hull:2009mi, Hull:2009zb, Hohm:2010jy, Hohm:2010pp}. 
The theory has since been subjected to a long series of elaborations and investigations, for example
\cite{
Hohm:2010xe,  Hohm:2011zr, Hohm:2011dv, Hohm:2011ex, Hohm:2011cp,
Hohm:2011nu,
Hohm:2013nja,
Jeon:2011vx, 
Jeon:2011sq, 
Jeon:2012kd, 
Jeon:2012hp, 
Jeon:2010rw, 
Jeon:2011cn, 
Hohm:2011si,
Hohm:2012mf, 
Berman:2013uda,
Grana:2012rr,
Aldazabal:2011nj, 
Berman:2013cli,
Berman:2011kg,
Geissbuhler:2011mx,
Geissbuhler:2013uka, 
Hohm:2012gk, Park:2013mpa, Berman:2014jba, Papadopoulos:2014mxa, 
Cederwall:2014kxa, Cederwall:2014opa,
Andriot:2012an, 
Berkeley:2014nza, Berman:2014jsa, 
Blumenhagen:2014gva,
Blumenhagen:2015zma 
}, for reviews see 
\cite{Aldazabal:2013sca,Berman:2013eva, Hohm:2013bwa}. 

In double field theory, the bosonic local symmetries are unified into generalised diffeomorphisms
acting on the doubled space. These give infinitesimal $O(D,D)$ transformations, whose finite
exponentiations in backgrounds with isometry give the usual notion of T-duality. The aim of this
paper is to seek the conserved quantities associated to generalised diffeomorphisms. This is the double
field theory version of constructing the Noether charge associated to diffeomorphisms in general
relativity. The latter, as we will briefly review in the next subsection, allows one to define a conserved
charge leading to the Komar integral. Our construction will hence produce a generalised Komar
integral, suitable for analysing the properties of solutions to double field theory. 

Note added: While we were finishing this paper, we became aware of the work by J.-H. Park, S.-J. Rey, W. Rim and Y. Sakatani, \cite{Park:2015bza},
which also studies conserved charges in double field theory.

\subsection{Recalling Noether's theorem}

The calculation that we will perform in this paper to obtain a conserved charge is essentially a
simple application of Noether's theorem, except to the perhaps slightly less familiar case that the
symmetry transformation involved is local. We will therefore now give a brief (and somewhat
schematic) review of the ideas we will be using, in order that it is clear what is happening in the remainder of the
paper. For further discussion, see for instance the textbook \cite{Ortin:2004ms}, from which we have
adapted this short review. 

Consider for simplicity a Lagrangian density $L( \phi_A, \partial_i \phi_A )$ for some set of fields
$\phi_A$ (of course in the cases we are interested in there will be terms involving two derivatives
of the fields, but these can be incorporated as a total derivative term). Suppose there is a global symmetry transformation $\delta$ such that $\delta L = \partial_i
f^i$ for some $f^i$. The action is then invariant, up to a surface term. Now, one has
\be
\begin{split} 
\delta S & = \int \left( 
\left( \frac{\partial L}{\partial \phi_A} - \partial_i \frac{\partial L}{\partial ( \partial_i \phi_A )} \right) \delta \phi_A 
+ \partial_i \left( \frac{\partial L}{\partial( \partial_i \phi_A ) } \delta \phi_A \right)\right) \\
& = \int \partial_i f^i \,.
\end{split} 
\ee
One concludes that \emph{on-shell}, when $\frac{\delta S}{\delta \phi} \equiv 
\frac{\partial L}{\partial \phi_A} - \partial_i \frac{\partial L}{\partial ( \partial_i \phi_A )}
 = 0$,
the current
\be
J_1^i = 
\frac{\partial L}{\partial ( \partial_i \phi_A )} \delta \phi_A 
- f^i 
\ee
is conserved. 

Now suppose instead we have a local symmetry transformation, so that $\delta \phi_A$ contains
derivatives of the symmetry parameters $\xi^I$. Integrating these by parts we obtain from the above
an identity of the schematic form
\be
0 = \int \left( \xi^I D_I \frac{\delta S}{\delta \phi}  
+ \partial_i J_2^i
\right) \,,
\ee
where $D_I$ represent some specific derivative operators acting on the equations of motion.
The current $J_2^i$ depends on the fields and the symmetry parameters. If one chooses the latter such
that $J_2^i$ vanishes on the boundary, then the terms proportional to $\xi^I$ give off-shell
identities which are the usual gauge or Bianchi identities. As these hold identically off-shell, one then also
obtains that the current $J_2^i$ is conserved \emph{off-shell}, $\partial_i J_2^i = 0$. (As a result, one
can always write $J_2^i = \partial_j J^{ij}$ with $J^{ij}$ antisymmetric.)

Note that one obtains in this way conserved currents which depend on the symmetry parameters
$\xi^I$. Thus, there seem to be an infinite number of such currents. However, when one seeks to
construct conserved \emph{charges} using these currents one has to consider integrals. These
integrals will not converge for arbitrary $\xi^I$, but only for certain parameters, which may then
often be associated with particular symmetry transformations of vacuum configurations. 

If one carries out this procedure for pure general relativity (in $D$ dimensions), one obtains a
current
\be
J^i = \sqrt{|g|} \nabla_j J^{ij} \quad , \quad
J^{ij} = 2 \nabla^{[j} \xi^{i]} \,.
\ee
From this we have a charge, integrating over some spatial hypersurface $\Sigma$,
\be
Q = \int_\Sigma d\Sigma_i J^i = \int_{\partial \Sigma} d \Sigma_{i j} 2 \nabla^j\xi^i \,,
\ee
which gives the Komar integral. Choosing $\xi^i$ to be timelike, for instance, corresponding to a
Killing vector $\xi = \partial_t$ in some coordinates, then this measures the Komar mass of a
background. 

One aspect of the application of this method to double field theory which is worth mentioning is the existence of the section condition constraint. Consistency of the theory leads one to impose the constraints $\eta^{MN} \partial_M \partial_N \mathcal{O} = 0$ and $\eta^{MN} \partial_M \mathcal{O}_1 \partial_N \mathcal{O}_2 = 0$ on all fields and gauge parameters\footnote{Note though that an alternative is provided by using the Scherk-Schwarz method, which we discuss in section \ref{se:SS}.}. These constraints are termed the section condition, and the effect is to reduce the coordinate dependence to a physical $D$-dimensional subset of the $2D$ doubled coordinates. 

A consequence of this is that the action of double field theory is invariant under generalised diffeomorphisms only up to terms which vanish upon imposing the section condition. Despite this, one is still able to extract and define a conserved current. This current, however, is only itself conserved up to terms which vanish by the section condition, that is $\partial_M J^M = \eta^{MN} \partial_M ( \dots) \partial_N ( \dots )$ = 0. This is consistent with the imposition of the section condition by hand. One might wonder whether there exists a formulation of the theory in which it arises naturally perhaps as a consequence of equations of motion, but in the present state we must take the section condition to be an additional constraint which we will require even on off-shell configurations. This ensures the off-shell Noether current is still a conserved current. 

\subsection{Outline of this paper}

Our starting point in section \ref{sec:2} will be the original action for double field theory. This is written
in terms of a Lagrangian density which transforms as a scalar of weight one under generalised
diffeomorphism (up to terms which vanish by the section condition). We then
compute the variation of the action under generalised diffemorphisms. Integrating by parts to remove
all derivatives from $\xi^M$, one finds the Bianchi identities of double field theory - as
previously obtained by this method in a number of papers \cite{Siegel:1993th, Kwak:2010ew, Hohm:2010xe, Hohm:2011si} - and an identically conserved
current. We write down this current in section \ref{sec:3}.

In section \ref{sec:4}, we test the practical uses of our current to define conserved charges by analysing the
fundamental string/pp-wave T-duality chain, which was the subject of the paper
\cite{Berkeley:2014nza} analysing this solution within double field theory.

In section \ref{sec:5}, we consider the effect of modifying the action of double field theory with the
addition of extra terms. We write down the extensions to the current that arise when adding total
derivatives, the so-called Scherk-Schwarz term, and also when the Ramond-Ramond sector is included.

We conclude in section \ref{sec:6} with some discussion about future work. We also include an appendix
collecting many useful facts about the geometric formalisms used.

\section{The action of double field theory and its variation}
\label{sec:2}

\subsection{A short summary of double field theory}

Double field theory is defined on a $2D$-dimensional space equipped with a background constant $O(D,D)$ structure, 
\be
\eta_{MN} = \begin{pmatrix} 0 & \mathbb{I} \\ \mathbb{I} & 0 \end{pmatrix} \,, 
\ee 
with inverse $\eta^{MN}$. The dynamical metric is known as the generalised metric, denoted
$\gM_{MN}$, which parametrises the coset $O(D,D)/O(D)\times O(D)$.\footnote{More properly,
$O(D,D)/O(D-1,1) \times O(D-1,1)$ if our $D$ directions include the time direction, as will be the
case for us in section \ref{sec:4}.} Hence it obeys the condition
$\gM_{MP} \gM_{NQ} \eta^{PQ} = \eta_{MN}$. If we introduce the conventional shorthand $S_M^N \equiv
\gM_{MP} \eta^{PN} \equiv \eta_{MP} \gM^{PN}$ this condition is equivalent to $S_M^P
S_P^N = \delta_M^N$. This in turn allows one to introduce projectors
\be
P_M^N = \frac{1}{2} ( \delta_M^N  - S_M^N ) \quad , \quad  
\bar{P}_M^N = \frac{1}{2} ( \delta_M^N + S_M^N ) \,,
\label{eq:theprojectors}
\ee
which play an important role. In this paper, we follow the usual convention of raising and lowering indices on these projectors using the flat $O(D,D)$ structure, rather than the generalised metric. 

One may also decompose the metric in terms of a generalised vielbein, $E^\alpha{}_M$, such that
$\gM_{MN} = \gM_{\alpha \beta} E^\alpha{}_ME^\beta{}_N$, where $\gM_{\alpha \beta}$ is the flat
generalised metric which we may be take to be given by $\gM_{\alpha \beta} = \mathrm{diag} ( h_{\mu
\nu}, h^{\mu \nu} )$ where $h_{\mu \nu}$ is a flat metric (Euclidean or Lorentzian depending on
whether our $D$ directions include time). This flat
generalised metric is preserved by local transformations in $O(D)\times O(D)$ which also act on the
flat index of the vielbein. 

The local symmetries of the doubled space are generalised diffeomorphisms, acting on a generalised vector $V^M$ of weight $w$ as
\be
\delta_\xi V^M = \xi^N \partial_N V^M - V^N \partial_N \xi^M + \eta^{MN} \eta_{PQ} \partial_N \xi^P V^Q + w \partial_N \xi^N V^M \,.
\label{eq:gendiffeo}
\ee
The generalised metric is a tensor of rank two under such transformations. We also introduce a
generalised dilaton, $d$, such that $e^{-2d}$ is a scalar of weight one. This will be used as an
integration measure. The definition \eqref{eq:gendiffeo} provides a generalised Lie derivative which can be extended to arbitrary tensors in the usual way, and is such that the $O(D,D)$ structure $\eta_{MN}$ is invariant: $\delta_\xi \eta_{MN} = 0$. 

The algebra of generalised diffeomorphisms does not close unless one imposes specific constraints. The simplest choice is to require the section condition 
\be
\eta^{MN} \partial_M f \partial_N g = 0 
\quad , \quad
\eta^{MN} \partial_M \partial_N f = 0 \,,
\ee
where $f,g$ may be any fields or gauge parameters of the theory. Alternatively, one may require that
all tensors in the theory can be written in the form of a Scherk-Schwarz ansatz \cite{Grana:2012rr},
as we describe in section \ref{sec:5}.

The original action of double field theory may be written as 
\cite{
Hull:2009mi,
Hull:2009zb, 
Hohm:2010jy,
Hohm:2010pp}
\be
S_{DFT} = \int dx d\tilde x e^{-2d} \mathcal{R} \,,
\label{eq:SDFT}
\ee
where we have 
\be
\begin{split}
\mathcal{R} & = \frac{1}{8} \gM^{M N} \partial_M  \gM^{PQ} \partial_N  \gM_{PQ}
- \frac{1}{2}  \gM^{MN} \partial_M  \gM^{PQ} \partial_P  \gM_{QN}\\
& + 4 \partial_M  \gM^{MN} \partial_N d 
- 4  \gM^{MN} \partial_M d \partial_N d\\
& - \partial_M \partial_N  \gM^{MN} + 4  \gM^{MN} \partial_M \partial_N d\,,
\end{split}
\label{eq:RDFT}
\ee
which is a scalar under generalised diffeomorphisms, up to terms which vanish by the section
condition. (We will frequently shorten $\int dx d\tilde x$ to a mere $\int$ to save space in subsequent expressions.)

\subsection{Connections for double field theory} 

In this paper, we will make use of two different covariant derivatives.

The first is the semi-covariant/semi-determined connection adopted in 
\cite{Jeon:2010rw, Jeon:2011cn, Hohm:2011si}. We denote the connection coefficients by
$\Gamma_{MN}{}^P$ and the covariant derivative operator by $\nabla_M$. This covariant derivative
annihilates the generalised metric, the generalised dilaton and the $O(D,D)$ structure. In addition,
in analogy to the Levi-Civita connection in general relativity, it has vanishing generalised
torsion. The latter condition means that generalised diffeomorphisms may be interchangeably written
in terms of $\nabla_M$ or $\partial_M$. The explicit expression for the generalised torsion is
\be
\tau_{MN}{}^P (\Gamma) = \Gamma_{MN}{}^P - \Gamma_{NM}{}^P + \eta^{PQ} \eta_{NR} \Gamma_{QM}{}^R \,.
\ee
If one lowers the upper index with $\eta_{MN}$ then this is totally antisymmetric, owing to the
condition $\Gamma_{MN}{}^P \eta_{PQ} = - \Gamma_{MQ}{}^P \eta_{PN}$, which must be true for any
connection which is compatible with the $O(D,D)$ structure. 

In general relativity, the Levi-Civita is uniquely determined by the requirements of metric
compatibility and vanishing torsion. The connection $\Gamma_{MN}{}^P$ in double field theory is,
however, not uniquely fixed by the above conditions. In
\cite{Hohm:2011si}, the connection is explicitly determined in terms of $\gM_{MN}$ and $d$ up to
certain undetermined components. In \cite{Jeon:2010rw, Jeon:2011cn}, one effectively drops these
undetermined components, giving a derivative which is only ``semi-covariant''. One may still obtain
covariant expressions by acting with certain combinations of the projectors
\eqref{eq:theprojectors}. The effect of these projections would be to also remove the undetermined
components present in the connection as used in \cite{Hohm:2011si}. 

Regardless of these issues, one can still define generalised versions of the Riemann tensor and thence by various contractions the Ricci tensor and scalar: the latter is found (with all undetermined or non-covariant components dropping out) to give exactly the scalar \eqref{eq:RDFT}. We give many explicit details of these constructions in the appendix. 

The second connection we will use appears rather simpler. This is the Weitzenb\"ock connection
\cite{Berman:2013uda}. We denote this by $\Omega_{MN}{}^P$, and can immediately give the explicit
expression for the connection coefficients as $\Omega_{MN}{}^P = E_\alpha{}^P
\partial_M E^\alpha{}_N$, where $E^\alpha{}_M$ is the generalised vielbein. We denote the
corresponding covariant derivative by $D_M$. This connection has vanishing generalised Riemann tensor,
but non-vanishing generalised torsion. 

The generalised Ricci scalar can then be recovered by seeking a scalar quadratic in this torsion and the covariant derivative of the dilaton (which is not zero for this connection) which is invariant under the local $O(D) \times O(D)$ transformations acting on the vielbein. This requirement gives a unique action which agrees with $\mathcal{R}$ up to a term which vanishes by the section condition - this term is however necessary to include when performing a Scherk-Schwarz reduction of the action. 

Both of these connections have uses within double field theory. In this paper, we will see that the
first connection appears very naturally in the conserved current associated to generalised
diffeomorphisms, in a manner exactly analogous to the Levi-Civita connection in the same situation
in general relativity. The Weitzenb\"ock connection, on the other hand, is known to be useful in
studying Scherk-Schwarz compactifications \cite{Geissbuhler:2011mx, Berman:2013uda}. We wish to also stress in this
paper that its generalised torsion - the primary geometric object associated to it - also defines a
notion of charge, closely associated to the idea of geometric and non-geometric flux. One may wish
to think of the connection $\Gamma_{MN}{}^P$ as being of use when focusing on aspects of double field
theory which appear ``gravitational'' in nature, while the connection $\Omega_{MN}{}^P$ is more
closely related to situations which resemble that of two-form gauge field. Indeed, the totally
antisymmetric generalised torsion $\tau_{MNP}$ is the direct generalisation within double field
theory of the field strength $H_3 = d B_2$. 

\subsection{Variation of the action} 

Although it would be convenient to work within a particular geometric framework, it remains
straightforward at this point to vary the above action directly:
\be
\begin{split}
\delta S_{DFT} & = 
\int e^{-2d} \left( - 2 \delta d \mathcal{R}  + \delta \gM^{MN} K_{MN} \right) \\ 
& + 
\int \partial_P \Big( 
e^{-2d} \delta \gM^{MN} 
\left(
\frac{1}{4} \gM^{PQ} \partial_Q \gM_{MN}  
- \frac{1}{2} \gM^{PQ} \partial_M \gM_{QN}
- \frac{1}{2} \gM^{KQ} \eta_{QM} \partial_K \gM^{PL} \eta_{LN} 
\right)
\\ 
& \qquad + 4 e^{-2d} \gM^{PN} \partial_N \delta d 
- e^{-2d} ( \partial_N - 2 \partial_N d ) \delta \gM^{PN} 
\Big) \,,
\end{split} 
\ee
where
\be
\begin{split}
K_{MN} & = 
\frac{1}{8} \partial_M \gM^{KL} \partial_N \gM_{KL} 
- \frac{1}{2} \partial_{(M|} \gM^{KL} \partial_K \gM_{|N)L}
+ 2 \partial_M \partial_N d 
\\ & 
+ \left( \partial_P - 2 \partial_P d\right) 
\left( 
-\frac{1}{4} \gM^{PQ} \partial_Q \gM_{MN} 
+ \frac{1}{2} \gM^{PQ} \partial_{(M} \gM_{N)Q} 
+ \frac{1}{2} \gM^{KQ} \eta_{Q(M} \partial_K \gM^{PL} \eta_{N)L} \right) \,,
\end{split}
\ee
As the generalised metric
is constrained to parametrise the coset $O(D,D)/O(D) \times O(D)$ its true
equation of motion is not $K_{MN} = 0$ but given in terms of the generalised Ricci tensor $\mathcal{R}_{MN}$
which is a projection of $K_{MN}$:
\be
\begin{split}
\mathcal{R}_{MN} 
& = \left( 
P_M^P \bar{P}_N^Q 
+ \bar P_M^P {P}_N^Q 
\right) K_{PQ} 
\\
& = \frac{1}{2} \left( \delta_M^P \delta_N^Q - S_M^P S_N^Q \right) K_{PQ}  \,.
\end{split}
\ee
We now specialise to the variation of the action under generalised diffeomorphisms. Acting on the metric and dilaton, these are given by
\be
\begin{split} 
\delta_\xi \gM^{MN} & = \xi^P \partial_P \gM^{MN} 
- \gM^{PN} \partial_P \xi^M + \eta^{MP} \eta_{RS} \partial_P \xi^R \gM^{SN} 
\\ & \qquad - \gM^{MP} \partial_P \xi^N + \eta^{NP} \eta_{RS} \partial_P \xi^R \gM^{MS} \\
& = \xi^P \partial_P \gM^{MN} 
- \gM^{PQ} \partial_P \xi^R \left( \delta_R^M \delta_Q^N - S_R^M S_Q^N \right) \\ & \qquad 
- \gM^{PQ} \partial_P \xi^R \left( \delta_R^N \delta_Q^M - S_R^N S_Q^M \right) \,,
\end{split} 
\label{eq:gendiffeoH}
\ee
and
\be
\delta_\xi d = \xi^P \partial_P d - \frac{1}{2} \partial_P \xi^P\,.
\label{eq:gendiffeod}
\ee
Using this,
\be
\begin{split}
\delta_\xi S_{DFT} & = 
\int e^{-2d} \xi^P \Big(
- \partial_P \mathcal{R} + \mathcal{R}_{MN} \partial_P \gM^{MN} 
+ 4 ( \partial_R - 2 \partial_R d) \gM^{RQ} \mathcal{R}_{PQ} 
\Big)
\\
& + 
\int \partial_P \Big( e^{-2d}
\Big[ 
-4  \xi^R \gM^{PQ} \mathcal{R}_{RQ} 
\\ & \qquad + 
 \delta_\xi \gM^{MN}  
\left(
\frac{1}{4} \gM^{PQ} \partial_Q \gM_{MN}  
- \frac{1}{2} \gM^{PQ} \partial_M \gM_{QN}
- \frac{1}{2} \gM^{KQ} \eta_{QM} \partial_K \gM^{PL} \eta_{LN} 
\right)
\\ 
& \qquad + 4  \gM^{PN} \partial_N \delta_\xi d 
-  ( \partial_N - 2 \partial_N d ) \delta_\xi \gM^{PN} 
\Big]
\Big) 
\\ & + 
\int \partial_P \left( e^{-2d} \mathcal{R} \xi^P \right) 
\,,
\end{split} 
\ee
and we know however that we must have $\delta_\xi S_{DFT} = 
\int \partial_P \left( e^{-2d} \mathcal{R} \xi^P \right)$, so that we can extract the identity
\be
\begin{split}
0 & = 
\int e^{-2d} \xi^P \Big(
- \partial_P \mathcal{R} + \mathcal{R}_{MN} \partial_P \gM^{MN} 
+ 4 ( \partial_R - 2 \partial_R d) \gM^{RQ} \mathcal{R}_{PQ}
\Big)
\\
& + 
\int \partial_P \Big( e^{-2d}
\Big[ 
-4 \xi^R \gM^{PQ} \mathcal{R}_{RQ}  
\\ & \qquad + 
 \delta_\xi \gM^{MN}  
\left(
\frac{1}{4} \gM^{PQ} \partial_Q \gM_{MN}  
- \frac{1}{2} \gM^{PQ} \partial_M \gM_{QN}
- \frac{1}{2} \gM^{KQ} \eta_{QM} \partial_K \gM^{PL} \eta_{LN} 
\right) 
\\ 
& \qquad + 4  \gM^{PN} \partial_N \delta_\xi d 
-  ( \partial_N - 2 \partial_N d ) \delta_\xi \gM^{PN} 
\Big] \Big) \,.
\end{split} 
\label{eq:VaryResult0}
\ee
The first line of this should give Bianchi identities, while the rest defines a conserved current. 

\subsection{Bianchi identities}

The Bianchi identity appears here in the form 
\be
- \partial_P \mathcal{R}  + \partial_P \gM^{MN} \mathcal{R}_{MN} 
+ 4 ( \partial_R - 2 \partial_R d ) \gM^{RQ} \mathcal{R}_{PR} = 0 \,.
\ee
This can be written in covariantly as 
\be
- \nabla_P \mathcal{R} + 4 \gM^{MN} ( \nabla_M - 2 \nabla_M d) \mathcal{R}_{NP} + 2 \gM^{MN} \tau_{MP}{}^Q
  \mathcal{R}_{QN} = 0
  \,.
\ee
We now apply $\nabla_M d = 0$, $\tau_{MN}{}^P = 0$ and rewrite $\gM^{MN} = \bar{P}^{MN} - P^{MN}$ to
find that this is 
\be
- P_P^Q \nabla_Q \mathcal{R} + 4 \bar{P}^{MN} \nabla_M \mathcal{R}_{NP} 
- \bar{P}_P^Q \nabla_Q \mathcal{R}- 4 P^{MN} \nabla_M \mathcal{R}_{NP} = 0
\ee
which is just the sum of the Bianchi identities of \cite{Hohm:2011si}, see appendix
\ref{app:Rbi}.

\section{The current}
\label{sec:3}

\subsection{The current}

The easiest way to evaluate the current is to rewrite in terms of geometric quantities. We can take
the total derivative term in \eqref{eq:VaryResult0} and write out the generalised diffeomorphism terms using the covariant derivative $\nabla$ in place of partial derivatives (as this connection is generalised torsion free). In addition it is simple to replace
the terms involving partial derivatives of the metric with connections, as the covariant derivative
annihilates the generalised metric. Then one can tidy up using the contracted generalised Ricci
identity, \eqref{eq:contractedRicci}. Thus one finds 
\be
\begin{split}
J^M & = \nabla_N \left( 4 e^{-2d} \nabla_Q \xi^P \left( P_P^{[M} \bar{P}^{N]Q} - \bar{P}_P^{[M}
P^{N]Q} \right) \right)
\\ & - 2 \nabla_N \left( \eta^{MN} e^{-2d} S_P^Q \nabla_Q \xi^P \right) \,. \\
\end{split}
\label{eq:thecurrent} 
\ee
Although technically this covariant derivative is either semi-determined or semi-covariant, the presence of the projectors ensures that this form of the current is indeed fully
determined and covariant as it must be. 

The current can be written as 
\be
\begin{split}
J^M & = \nabla_N J^{MN} - 2 \nabla_N \left( \eta^{MN} e^{-2d} S_P^Q \nabla_Q \xi^P \right)  \\
 & = 
\partial_N J^{MN} + \frac{1}{4} \eta^{MN} \partial_N \gM_{PK} S_L^P J^{KL} 
- 2 \eta^{MN} e^{-2d} \partial_N \left( \partial_Q - 2 \partial_Q d\right) \left( S^Q_P \xi^P \right) \,,
\end{split} 
\label{eq:Jsimpler}
\ee
where we define the antisymmetric tensor of weight one
$J^{MN}$ via 
\be
\begin{split}
J^{MN} e^{2d} & \equiv  4 \nabla_Q \xi^P \left( P_P^{[M} \bar{P}^{N]Q} - \bar{P}_P^{[M} P^{N]Q}\right) 
\\ & =  
2 \partial_P \xi^{[M} \gM^{N]P} - 2 \partial_Q \xi^P S_P^{[M} \eta^{N]Q} \\ & 
\quad + \xi^P \left( 2 \eta^{Q[M} \partial_Q S^{N]}_P +2 \gM_{RP} \gM^{Q[M} \partial_Q \gM^{N] R}
-  S^Q_P S_R^{[M} \partial_Q \gM^{N]R} 
\right) \,.
\end{split} 
\label{eq:JMN}
\ee
Note that we have 
\be
\left( P_R^M \bar{P}_S^N + \bar{P}_R^M P_S^N \right) J^{RS} 
= J^{MN} \,.
\label{eq:Jproj}
\ee
which can be used to show  
\be
\nabla_N J^{MN} = \partial_N J^{MN} + \frac{1}{4} \eta^{MN} \partial_N \gM_{PK} S_L^P J^{KL} \,.
\ee
We have also used
\be
\nabla_Q ( S^Q_P \xi^P )  = ( \partial_Q - 2 \partial_Q d ) S^Q_P \xi^P \,.
\ee
The form \eqref{eq:Jsimpler} makes the conservation law $\partial_M J^M = 0$ manifest by the antisymmetry of $J^{MN}$, up to the section condition. It is interesting to note that the general form of an identically conserved current in double field theory is 
\be
J^M = \partial_N J^{MN} + \varphi^M \,,
\ee
with $J^{MN}$ antisymmetric and $\varphi^M \sim \phi^\prime \eta^{MN} \partial_N \phi$ for some (possibly index-carrying) fields $\phi, \phi^\prime$ such that $\partial_M \varphi^M = 0$ by the section condition: we will encounter the same general form in the conserved currents we calculate in section \ref{sec:5}. 

\subsection{Different forms of the current}

\subsubsection{Generalised Killing vector}

Suppose now that $\xi^M$ is a generalised Killing vector. Then $\delta_\xi \gM^{MN} =0$ and
$\delta_\xi d= 0$, which respectively imply the generalised Killing equations
\be
\left( 
P_K^M \bar{P}_L^N 
+ 
\bar{P}_K^M P_L^N 
\right) 
\nabla^{(K} \xi^{L)} = 0 \,,
\ee
(here the index on the covariant derivative has been raised with the generalised metric) and
\be
\nabla_M \xi^M = 0 \,.
\ee
If we refer to the original identity
\eqref{eq:VaryResult0} resulting from the variation, we see immediately that the current becomes
\be
J^M = - 4 e^{-2d} \xi^P \gM^{MN} \mathcal{R}_{PN}  \,.
\label{eq:KillingJ}
\ee
Something very similar is true in general relativity, where one finds the current associated to a Killing vector is similarly given by the contraction of the Killing vector and the Ricci tensor. 
Note that this means that on-shell, in the absence of sources, the current will vanish. The solution we study in section \ref{sec:4} is sourced, and so we will obtain a non-trivial current there. In general, one should also take into account boundary terms \cite{Berman:2011kg} which modify the current \cite{Park:2015bza, Naseer:2015fba}. 

One can also check, either by starting from \eqref{eq:KillingJ} and using the generalised Ricci
identity \eqref{eq:contractedRicci} or by starting from the expression \eqref{eq:thecurrent} for the
current and using the generalised Killing equations, that this is equivalent to
\be
\begin{split} 
J^M & = 4 \nabla_N \left( e^{-2d} \nabla^Q \xi^P \left[ P_P^M \bar{P}_Q^N + \bar{P}_P^M P_Q^N
\right] \right) 
- 2 \nabla_N \left( e^{-2d} \eta^{MN} S_Q^P \nabla_P \xi^Q \right) \\
 & = - 2 \nabla_N \left( e^{-2d}\left[ \nabla^M \xi^N - S^M_P S^N_Q \nabla^P \xi^Q \right] \right) 
 - 2 \nabla_N \left( e^{-2d} \eta^{MN} S_Q^P \nabla_P \xi^Q \right) \,.
\end{split} 
\ee

\subsubsection{The Weitzenb\"ock connection}

One can also rewrite the above form of the current in terms of the Weitzenb\"ock connection,
$\Omega_{MN}{}^P$, which has non-vanishing generalised torsion, $\tau_{MN}{}^P$. One has
\be
\begin{split}
J^{MN} e^{2d} & = 2 D_P \xi^{[M} \gM^{N]P} - 2 D_P \xi^Q S_Q^{[M} \eta^{N]P}
\\ & 
+ \xi^P \left( \eta^{MK} \eta^{NL} - \gM^{MK} \gM^{NL} \right) \gM_{PQ} \tau_{KL}{}^Q 
\,.
\end{split}
\label{eq:JWeitz}
\ee
and
\be
\nabla_N J^{MN} = D_N J^{MN} + \frac{1}{2} \tau_{KL}{}^M J^{KL} \,.
\ee
In addition,
\be
\nabla_Q ( S^Q_P \xi^P) = D_Q ( S^Q_P \xi^P) - 2 S^Q_P \xi^P D_Q d \,.
\ee
Thus, the current is
\be
J^M = D_N J^{MN} + \frac{1}{2} \tau_{KL}{}^M J^{KL} 
- 2 e^{-2d} \eta^{MN} D_N \left( D_Q (S^Q_P \xi^P ) - 2 S^Q_P \xi^P D_Q d \right) \,,
\ee
or
\be
J^M = \partial_N J^{MN} 
+ \frac{1}{2} \eta^{MN} \eta_{KL} \Omega_{NP}{}^K J^{PL} 
- 2 e^{-2d} \eta^{MN} \partial_N ( \partial_Q - 2 \partial_Q d ) \left( S^Q_P \xi^P \right)
  \,.
\label{eq:currentSimpleWeitz}
\ee

\section{Analysis and applications}
\label{sec:4}

\subsection{Reduction to spacetime}

To understand what our current contains, we can explicitly evaluate the components of the current in terms of the usual physical fields
and coordinate dependence. We take the usual section $\tilde \partial^i = 0$ and the usual parametrisation of the generalised
metric in terms of a spacetime metric and $B$-field,
\be
\gM_{MN} = \begin{pmatrix} 
g_{ij} - B_{ik} g^{kl} B_{lj} & B_{ik} g^{kj} \\
 - g^{ik} B_{kj} & g^{ij} 
\end{pmatrix} 
\,.
\ee
Then, one can show firstly that
\be
J^{ij} = \sqrt{|g|} e^{-2\phi} 
\left(
\nabla^j \xi^i - \nabla^i \xi^j - ( \lambda_k +\xi^p B_{pk} ) H^{ijk}
\right)
\,,
\ee
\be
J_i{}^j = \sqrt{|g|}e^{-2\phi} \left( \nabla^j (\lambda_i + \xi^p B_{pi} ) + \nabla_i ( \lambda^j +
\xi^p B_p{}^j ) \right) 
+ B_{ip} J^{pj} \,,
\ee
\be
 J_{ij} = - ( g_{ik} g_{jl} + B_{ik} B_{jl} ) J^{kl} + B_{ik} J_j{}^k - B_{jk} J_i{}^k \,.
\label{eq:Jijusual}
\ee
Here $\nabla_i$ is the usual spacetime Levi-Civita connection. Observe that equation
\eqref{eq:Jijusual} is a consequence of the projection identity \eqref{eq:Jproj}.

From this, one obtains that
\be
\nabla_N J^{iN} = \partial_j J^{ij} 
\ee
and
\be
\nabla_N J_{i}{}^N = \nabla_j \tilde J_i{}^{j} + \frac{1}{2} H_{imn} J^{mn} + B_{ip} \partial_j
J^{pj} \,,
\ee
where $\tilde J_i{}^j = J_i{}^j - B_{ik} J^{kj}$. 

Hence we get for the ``physical'' components of the current
\be
J^i = \partial_j \left( \sqrt{|g|} e^{-2\phi} 
\left[ \nabla^j \xi^i - \nabla^i \xi^j - ( \lambda_k +\xi^p B_{pk} ) H^{ijk}\right] \right) \,.
\label{eq:Jiphys}
\ee
We could replace the partial derivative with a
covariant one here - the connection terms vanish owing to the antisymmetry of $J^{ij}$ and the fact
that it has weight one. Meanwhile for the components with lower indices, we find
\be
\begin{split} 
J_i&  = \nabla_j \left( \sqrt{|g|} e^{-2\phi} \left[
\nabla^j (\lambda_i + \xi^p B_{pi} ) + \nabla_i ( \lambda^j + \xi^p B_p{}^j )\right] \right)
\\ & + \frac{1}{2}\sqrt{|g|}e^{-2\phi} H_{ijk} \left( 
\nabla^k \xi^j - \nabla^j \xi^k - ( \lambda_l +\xi^p B_{pl} ) H^{jkl}
\right) 
\\ & 
- 2 \sqrt{|g|} e^{-2\phi} \nabla_i \left( ( \nabla_j - 2 \nabla_j \phi) ( \lambda^j + \xi^p B_p{}^k ) \right) 
\\ & 
+ B_{ip} J^p \,.
\label{eq:Jinonphys}
\end{split} 
\ee 
Looking at the physical part of the current, equation \eqref{eq:Jiphys}, we see that setting both the dilaton and $B$-field to zero, we obtain exactly the expected conserved current
associated to diffeomorphisms that is obtained in general relativity. The resulting conserved charge
that one would define using this is the Komar integral. 

The part of the current \eqref{eq:Jiphys} involving $\lambda_k$ is the conserved current associated to gauge
transformations of the $B$-field. The resulting charge is just the electric charge of this field.
The additional piece involving $\xi^p B_{pk}$ appears to ensure this component of the current is
invariant under gauge transformations (as under a generalised diffeomorphism $\lambda_k$ itself
transforms).

\subsection{Reduction to spacetime with a bivector}

We should observe that the current $J^i$ we obtained above could have been derived directly from the action of the NSNS
sector of supergravity, without any reference to double field theory, by varying with respect to
both diffeomorphisms and gauge transformations. Given that the components $\tilde J_i$ do not appear
to have an obvious interpretation in spacetime, it may seem more natural to have done so. 

However, having access to a T-duality covariant form of the current allows one to explore other interesting scenarios. In this subsection, we will consider an alternative parametrisation of the generalised metric, in which one uses a bivector $\beta^{ij}$ in place of the $B$-field:
\be
\gM_{MN} = \begin{pmatrix}
g_{ij} & - g_{ik} \beta^{kj} \\
\beta^{ik} g_{kj} & g^{ij} - \beta^{ik} g_{kl} \beta^{lj} 
\end{pmatrix} \,.
\label{eq:genmetbeta}
\ee
This parametrisation is useful in non-geometric situations, in which the metric and $B$-field are not globally well-defined. The reduction of the double field theory action gives a complicated action which is closely related to that 
of ``$\beta$-supergravity'' \cite{
Andriot:2011uh, Andriot:2012wx, Andriot:2012an,
Andriot:2013xca
}, which was introduced as a means to study non-geometric fluxes in 10 dimensions. Note that we still impose the same section condition, $\tilde \partial^i = 0$ (one could also consider the entirely dual section $\partial_i = 0$, in which case for this bivector parametrisation the current components would take the same form as \eqref{eq:Jiphys} and \eqref{eq:Jinonphys} but with all index positions reversed, i.e. $B_{ij} \rightarrow \beta^{ij}$, $J^i \rightarrow J_i$, $J_i \rightarrow J^i$, etc.).  

It is tedious but straightforward to evaluate the current in this parametrisation. We find
\be
\begin{split} 
J_{ij} e^{2d} & = 2 \nabla_{[i} \tilde \xi_{j]} - 2 \nabla_k \lambda_{[i} \beta_{j]}{}^k
- \lambda^k \nabla_k \beta_{ij} 
- \tilde \xi_k R_{ij}{}^k \,,
\end{split}
\ee
\be
\begin{split}
J_i{}^j e^{2d} & = \nabla^j \lambda_i + \nabla_i \lambda^j
+ \beta_i{}^k \nabla_k \tilde\xi^j + \beta^{jk} \nabla_k \tilde \xi_i
+ \tilde \xi_k ( \nabla_i \beta^{jk} + \nabla^j \beta_i{}^k ) 
\\ & + \beta^{jl} J_{il} e^{2d} \,,
\end{split} 
\ee
%
and
\be
\begin{split}
J^{ij} e^{2d} & = 
- 2\nabla^{[i} \xi^{j]} - 2 \beta_k{}^p \beta^{k[i} \nabla_p \xi^{j]} 
\\ &  + 3 \lambda_k \nabla^{[i} \beta^{jk]}
+ \lambda_m \left( 2 \beta_k{}^p \beta^{k[i} \nabla_p \beta^{j]m} - \beta^{ik} \beta^{jl} \nabla^m
\beta_{kl} \right) 
\\ & 
+ \tilde \xi_m \left( - 2 \beta_l{}^{[i} \nabla^{j]} \beta^{lm} +  \beta_l{}^m \nabla^l \beta^{ij}
\right) 
-3\tilde \xi_m \beta^i{}_k \beta^j{}_l 
 \beta^{n[k} \nabla_n \beta^{lm]}
\,.
\end{split}
\ee
Here we introduced the shorthand
\be
\tilde \xi^j \equiv \xi^j + \lambda_k \beta^{kj} \,,
\ee
and have freely raised and lowered using the spacetime metric $g_{ij}$. 

Note that the projection identity \eqref{eq:Jproj} obeyed by
$J^{MN}$ implies that 
\be
J^{ij} = - ( g^{ik} g^{jl} + \beta^{ik}\beta^{jl} ) J_{kl} 
+ \beta^{ik} J_k{}^j - \beta^{jk} J_k{}^i \,.
\ee
One can then show that the current itself has components 
\be
J^i = \partial_j J^{ij} \,,
\ee
and
\be
\begin{split}
J_i & = \nabla_j \tilde J_i{}^j + \beta^{jk} \nabla_j J_{ik} + \frac{1}{2} \nabla_i \beta^{kl} J_{kl} +
\nabla_j \beta^{jk} J_{ik} 
\\ & - 2 \sqrt{|g|} e^{-2\phi} \nabla_i \left( (\nabla_k - 2 \nabla_k \phi ) \left( \lambda^k - \tilde \xi^l
  \beta_l{}^k \right) \right) \,.
\end{split}
\ee
Here, $\tilde J_i{}^j = J_i{}^j - \beta^{jk} J_{ik}$. There are some unexpected cancellations amongst the connection terms in the above expressions, so that 
\be
\beta^{jk} \nabla_j J_{ik} + \frac{1}{2} \nabla_i \beta^{kl} J_{kl}  +
\nabla_j \beta^{jk} J_{ik} 
= \beta^{jk} \partial_j J_{ik} + \frac{1}{2} \partial_i \beta^{kl} J_{kl} +
\partial_j \beta^{jk} J_{ik} \,,
\ee
and
\be
 3 \beta^{l[i} \nabla_l \beta^{jk]} = 3 \beta^{l[i} \partial_l \beta^{jk]} \,.
\ee
(The latter is essentially (part of) the $R$-flux.)

On general grounds, we expect that the current here gives an electric charge associated to the bivector field. The precise identification of this charge is perhaps not obvious as the form of the current is quite complicated. Note that the parametrisation of the generalised metric \eqref{eq:genmetbeta} implies unusual and non-linear transformations of the spacetime fields under gauge transformations (of the now defunct $B$-field):
\be
\begin{split} 
\delta_\lambda g_{ij} & = ( \partial_i \lambda_k - \partial_k \lambda_i ) \beta^{kl} g_{lj} 
+ ( \partial_j \lambda_k - \partial_k \lambda_j ) \beta^{kl} g_{li} \,, \\
\delta_\lambda \beta^{ij} & = ( g^{ik} g^{jl} - \beta^{ik} \beta^{jl} ) ( \partial_k \lambda_l - \partial_k \lambda_l ) \,.
\end{split}
\ee
Thus, the metric itself in this parametrisation transforms under these transformation. 

In practice, the only configurations involving a bivector that we will consider will be obtained by
acting with T-duality on solutions with a $B$-field. In such cases it will always be possible to
calculate the current in whichever parametrisation is easiest. However, we will keep in mind the
general idea that there may exist non-geometric configurations defined solely in terms of a
bivector, in which case one would have to deal directly with the above expressions (or at least the
expression for the current in terms of the components of the generalised metric, which may be
simpler. In general, one only needs to consider expressions in terms of spacetime fields when one is
interested in understanding the interpretation in spacetime itself). For now, we leave an
understanding of the physical nature of the above current to further study.

\subsection{Defining the charge}

The current we have found obeys a conservation equation 
\be
\partial_M J^M = 0 \,.
\ee
If we impose the section condition $\tilde\partial^i = 0$ then this reduces to
\be
\partial_i J^i = 0 \,.
\ee
From the point of view of a spacetime reduction, it seems therefore that only the $J^i$ components
of the current are relevant, giving a physical conserved current. Conventionally, given such a
current one would define a conserved charge by integrating the time component over a spatial
hypersurface, that is (up to some normalisation)
\be
Q = \int_{\Sigma_{D-1}} J^t \,.
\ee
(One could also write this covariantly in terms of a normal vector to the hypersurface,
$\int_{\Sigma_{D-1}} n_i J^i$.) 

In practice, we will adopt this definition of the charge in spacetime. One could also consider a more doubled version of the above, in which one would integrate also over an extra $D-1$ dimensional hypersurface, which after solving the section condition, would lie solely in the dual directions. This integration would give an extra volume factor, making a different normalisation of the current necessary. In either case, we may think of the charge as defining a generalised Komar integral. 

\subsection{Application to the fundamental string/pp-wave}

The fundamental string (F1) solution is T-dual to a pp-wave. These solutions were analysed in double
field theory in the paper \cite{Berkeley:2014nza}, and were shown to appear as a generalised pp-wave in the doubled space. To describe this solution, we split the $2D$ doubled coordinates
$X^M = ( t, z, y^\mu, \tilde t, \tilde z, \tilde y_\mu)$, with $\mu = 1, \dots, D-2$ (in practice
one can just take $D=10$ here, but we leave it general). The generalised metric can be read
off from the following generalised line element
\be
\begin{split} 
ds^2 & = ( H - 2) ( dt^2 - dz^2) 
+ 2 (H-1) (dt d\tilde z + d\tilde t dz ) 
\\ & - H ( d \tilde t\,{}^2 - d \tilde z^2 ) 
+ \delta_{\mu \nu} dy^\mu dy^\nu
+ \delta^{\mu \nu} d\tilde y_\mu d\tilde y_\nu \,,
\end{split}
\label{eq:DFTpp} 
\ee
where $H$ is a harmonic function of the $D-2$ coordinates $y^\mu$, $H = 1 + \frac{h}{r^{D-4}}$ with
$r = \sqrt{ \delta_{\mu \nu} y^\mu y^\nu}$. The generalised dilaton is constant. 
This solution has the form of a pp-wave travelling in the
$\tilde z$ direction. We will demonstrate below by constructing the conserved current for this
solution that it indeed carries a conserved charge which can be interpreted as momentum in this
direction, as conjectured in \cite{Berkeley:2014nza}.

By choosing the section
such that $(t,z,y^\mu)$ are the physical coordinates and decomposing the generalised metric, one
obtains the metric, $B$-field and dilaton of the fundamental string solution:
\be
\begin{split} 
ds^2 & = - H^{-1} dt^2 + H^{-1} dz^2 + \delta_{\mu \nu} dy^\mu dy^\nu \\
B & = ( 1 - H^{-1} ) dt \wedge dz \\
e^{-2\phi} & = H \,.
\end{split}
\label{eq:F1soln} 
\ee
The string itself lies in the $z$ direction.
If one instead chooses the section condition such that $(t,\tilde z, y^\mu)$ are the physical
coordinates, one obtains the pp-wave: 
\be
\begin{split}
ds^2 &  = - ( 2 - H ) dt^2 + 2 ( H - 1) dt d\tilde z + H d\tilde z^2 + \delta_{\mu \nu} dy^\mu
dy^\nu \,, \\ 
B & = 0 \,, \\
e^{-2\phi} & = 1 \,.
\end{split} 
\label{eq:ppwave}
\ee
These two solutions are related by a Buscher T-duality acting to interchange the $z$ and $\tilde z$
coordinates. 

One can now proceed to calculate the conserved current and charge. There are a number of ways to do so. In
order to have a check on our results, and to demonstrate a further subtlety relating to the choice
of section, we will proceed by calculating the $J^t$ component for both the fundamental string and
the pp-wave independently. As expected by T-duality, we will obtain the same result in either case,
but with a different interpretation. 

\subsubsection{Current from the F1 solution}

For the fundamental string, the field strength of the $B$-field has non-vanishing components $H_{\mu
t z} = H^{-1} \partial_\mu \log H$. The only
non-zero components of the Christoffel symbols are
\be
\Gamma_{tt}{}^\mu = - \Gamma_{zz}{}^\mu = - \frac{1}{2} H^{-2} \partial^\mu H 
\quad , \quad
\Gamma_{\mu z}{}^z = \Gamma_{\mu t}{}^t = - \frac{1}{2} H^{-1} \partial_\mu H \,.
\ee
The current is given by $J^i = \partial_j J^{ij}$ and the timelike components needed arise from 
\be
J^{t \mu} = \partial^\mu \xi^t + H \partial_t \xi^\mu - \partial^\mu H ( \xi^t + \lambda_z )  \,,
\ee
\be
J^{t z} = - H \partial_t \xi^z - H \partial_z \xi^t - \lambda_\mu \partial^\mu H \,.
\ee
Hence,
\be
\begin{split} 
J^t&  = \partial_\mu \left(  \partial^\mu \xi^t + H \partial_t \xi^\mu 
- \partial^\mu H ( \xi^t + \lambda_z ) \right) 
\\ &\quad + \partial_z \left(- H \partial_t \xi^z - H \partial_z \xi^t - \lambda_\mu \partial^\mu H 
\right) \,. 
\end{split} 
\label{eq:F1Jt}
\ee
If we specialise to the case where the gauge parameters are constant then
\be
J^t = - ( \xi^t + \lambda_z ) \partial_\mu \partial^\mu H \,.
\ee
Importantly, if we integrate this over a spatial hypersurface to define a charge, we do not get zero
as the derivatives of the harmonic function give a delta function at $r=0$. (We will carry out this
calculation more carefully below.) Thus we obtain charge
associated to constant $\xi^t$ - Killing vector generating translations in the time direction - and
constant $\lambda_z$. The former we interpret as mass. The latter may be better
interpreted in DFT. There $\lambda_z$ generates translations in the \emph{dual} $\tilde z$
direction. This resulting charge is then just the momentum in this dual direction, as expected. In
this frame it should be thought of as a winding charge. 

\subsubsection{Current from the pp-wave}

Suppose now that we directly calculate the current associated to the pp-wave. There is no
$B$-field, and the components of the Christoffel symbols are
\be
\Gamma_{tt}{}^\mu = \Gamma_{\tz \tz}{}^\mu = \Gamma_{t \tz}{}^\mu = - \frac{1}{2} \partial^\mu H \,,
\ee
\be
\Gamma_{\tz \mu}{}^{\tz} = \Gamma_{t\mu}{}^{\tz} = 
- 
\Gamma_{t\mu}{}^t = - \Gamma_{\tz \mu}{}^t = \frac{1}{2} \partial_\mu H \,.
\ee
One obtains
\be
\begin{split} 
J^{t} & = 
\partial_\mu \left( H ( \partial_t - \partial_{\tz} ) \xi^\mu 
+ \partial_{\tz} \xi^\mu + \partial^\mu \xi^t 
- \partial^\mu H ( \xi^t + \xi^{\tz} \right)
 \\ 
& \quad + \partial_{\tz} \left( 
H \partial_t \xi^{\tz} + ( 1 - H ) \partial_{\tz} \xi^{\tz} 
+ ( H-1) \partial_t \xi^t + ( 2 - H) \partial_{\tz} \xi^t 
\right) \,.
\end{split} 
\ee
At first glance, this does not appear to match the expression \eqref{eq:F1Jt}. This is because we
have implicitly and subtly used a different section condition choice. The derivatives with respect
to the coordinate $z$ in the F1 solution would be derivatives with respect to a dual coordinate in
the pp-wave solution. However, here we have evaluated the current in the natural section choice for
the the pp-wave, in which we have allowed dependence on its physical coordinate $\tz$. In order to
directly compare the two versions of the current we could then either return to the original,
T-duality invariant expression and include dual derivatives. Alternatively, we can take the simpler
approach of imposing the isometry condition $\partial_z = 0 = \partial_{\tz}$ on our gauge
parameters. Then, one has
\be
J^t_{F1} 
 = \partial_\mu \left(  \partial^\mu \xi^t + H \partial_t \xi^\mu - \partial^\mu H ( \xi^t + \lambda_z )
 \right) \,,
\ee
and
\be
J^t_{pp} = 
\partial_\mu \left( 
\partial^\mu \xi^t + +   \partial_t  \xi^\mu 
- \partial^\mu H ( \xi^t + \xi^{\tz} ) \right) 
\,.
\ee
These two expressions agree, as T-duality acting on the gauge parameter $\xi^M$ gives $\xi^{\tz} =
\lambda_z$. 

\subsubsection{Explicit calculation of the charge}

Let us now demonstrate in more detail that we do really get the expected charges. The mass of the fundamental string wrapped in the $z$ direction is expected to be $M_{F1} = \frac{R_z}{l_s^2}$ (where $l_s = \sqrt{\alpha^\prime}$ is the string lengthscale). The electric charge of the string is just its tension, $q_{F1} = \frac{1}{2\pi l_s^2}$. If we wrap this on the $z$ direction we pick up a factor of $2\pi R_z$, so that we have $q_{F1} = M_{F1}$. This is just the Bogomolny bound.  

We now commit ourselves to $D=10$ and define a properly normalised charge by
\be
Q = \frac{ e^{2\phi_0} }{16 \pi G_N^{(10)}} \int_{\Sigma} J^t \,,
\ee
where the ten dimensional Newton's constant is $G_N^{(10)} = 8 \pi^6 l_s^8 e^{2\phi_0}$ (we have
followed closely the conventions and definitions of the textbook \cite{Ortin:2004ms}) and
$e^{\phi_0}$ is the asymptotically value of the string coupling constant), which we will set to
$1$.\footnote{As we have passed already to a physical section, we have chosen to use the
conventional notation. In double field theory, we should instead write $e^{2d_0}$, where $d_0$ will
be the asymptotic value of the generalised dilaton. For spacetime metrics that are asymptotically
flat, this will agree with $\phi_0$. We implicitly make this assumption in defining our charges
here. Properly speaking, we should be more careful about our assumptions regarding the asymptotic
behaviour of the generalised metric. It will be important to bear this in mind in future use and
study of the conserved current.}
We are integrating over some spatial hypersurface $\Sigma$. 

For the fundamental string, the $z$ direction is assumed to be compact, and so we have
\be
\begin{split}
Q&  = - \frac{2\pi R_z}{16 \pi G_N^{(10)}} (\xi^t + \lambda_z) \int_{\Sigma_8} \partial_\mu \partial^\mu H \\
 & = - \frac{R_z}{8 G_N^{(10)}} (\xi^t +\lambda_z) \int_{ \partial \Sigma_8 = S^7_\infty} n_\mu \partial^\mu H \,,
\end{split}
\ee
using Stokes' theorem. Now, $\partial^\mu H = - 6 h x^\mu r^{-8}$, the normal is $n_\mu = x_\mu r^{-1}$ and as a result the powers of $r$ all cancel when we take into account the volume form for the integration. We are left with a simple integration over the unit sphere, and after substituting in for $G_N^{(10)}$ and using the fact that the constant appearing in the harmonic function is determined to be $h = ( 2 \pi l_s)^6 / ( 6 \mathrm{Vol}(S^7) )$ we obtain the beautifully simple result 
\be
Q = ( \xi^t + \lambda_z ) \frac{R_z}{l_s^2} \,,
\ee
which gives us exactly the expected values (taking separately $\xi^t = 1$, $\lambda_z = 1$, corresponding to the generalised Killing vectors $\partial_t$ and $\tilde \partial^z$). Hence with this normalisation, we obtain exactly the expected mass and electric charge for the fundamental string solution. The latter is interpreted in double field theory as momentum in the dual direction. 

After T-duality to the pp-wave, we can use the familiar relation $R_{\tilde z} = l_s^2 / R_z$ to see that we obtain in its frame
\be
Q = ( \xi^t + \xi^{\tz} ) \frac{1}{R_{\tz}} \,,
\label{eq:Qpp}
\ee
giving the expected result for the mass and momentum of a Kaluza-Klein state.\footnote{One might worry that the result of carrying out the integration directly in the case of the pp-wave should give $R_{\tz}/l_s^2$ rather than the above. However, if one treats carefully the duality keeping explicit track of the factors of the radius, one discovers that the dilaton picks up a constant shift $e^{\phi_0} \rightarrow e^{\phi_0} l_s / R_z$. This modifies $G_N^{(10)}$ for the pp-wave solution in exactly the right way so that one obtains the expected charge \eqref{eq:Qpp}.}

\subsubsection{Timelike T-duality}

There remains one additional isometry direction: $t$. In double field theory, making the naive
choice $(\tilde t, \tilde z, y^\mu)$ of the section condition is perfectly valid, and should give a solution which is timelike dual to the pp-wave and F1. 

Such a solution offers interesting connection with ``exotic branes''. 
While the fundamental string couples electrically to the $B$-field, the NS5 brane couples magnetically. T-duality maps the F1 to the pp-wave, which is charged electrically under the Kaluza-Klein graviphoton, and also maps the NS5 brane to the Kaluza-Klein monopole, which is coupled magnetically to the graviphoton. A further T-duality takes us from the Kaluza-Klein monopole to an exotic brane known as the $5_2^2$ \cite{deBoer:2010ud, deBoer:2012ma}. This brane is non-geometric - the metric and $B$-field are only globally defined up to a T-duality - but well-defined in double field theory. The solution can be more naturally expressed in terms of the bivector $\beta^{ij}$, and one can think of the $5_2^2$ as being magnetically charged under this field (thinking of $\beta^{ij}$ as a 0-form carrying two upper indices corresponding to special isometry directions, similar to the one-form Kaluza-Klein vector which carries an upper index corresponding to the special Kaluza-Klein direction). 

One is therefore led to wonder about whether there is an electric counterpart to the $5_2^2$, which inevitably would be generated by timelike duality \cite{Bergshoeff:2011se}. Indeed, such a solution has recently been written down in \cite{Sakatani:2014hba}, in the context of $\beta$-supergravity. It is also interesting to note that in \cite{Berman:2014hna} it is argued that the duality frame with coordinates $(\tilde t, \tz, y^\mu)$ is naturally selected as the appropriate description of the double field theory F1/pp-wave solution near the singularity at $r=0$. 

Picking this duality frame, or carrying out the T-duality on the pp-wave, one obtains the
configuration
\be
\begin{split}
ds^2 & = - \frac{1}{2-H} d\tilde t\,{}^2 + \frac{1}{2-H} d\tz^2 + \delta_{\mu \nu} dy^\mu dy^\nu \,,\\
B & = \frac{H-1}{2-H} d\tilde t \wedge d \tilde z \,,\\
e^{-2\phi} & = | 2-H|  \,.
\end{split} 
\ee
The metric, $B$-field and its field strength are now singular at $r =
h^{1/(D-4)}$. 
The generalised metric is not singular here, though. We can alternatively express the solution using
the bivector parametrisation (as is in fact sometimes necessary when considering timelike dualities
\cite{Malek:2013sp}), obtaining the configuration
\be
\begin{split}
ds^2 & = - H d\tilde t\,{}^2 + H d\tz^2  + \delta_{\mu \nu} dy^\mu dy^\nu \,,\\
\beta & = (1-H^{-1})  \partial_{\tilde t} \wedge \partial_{\tz}  \,,\\
e^{-2\phi} & = H^{-1} \,.
\end{split} 
\ee
The form of the bivector is essentially identical to that of the $B$-field of the fundamental string
solution. Thus one would like to think of it as an object which is electrically
coupled to $\beta$. We shall refer to it as $\widetilde{\mathrm{F}1}$. 

Let us now think about the conserved charges associated to this solution. As this background is obtained by a T-duality interchanging $t$ and $\tilde t$, the timelike component that is relevant for the charge is in fact $J^{\tilde t}_{\widetilde{F1}} = J_{t F1}$. One can calculate this component starting with the form of the solution above - it does not matter whether one uses the singular configuration in terms of the $B$-field or the more natural description using the bivector, although the former is actually simpler given the form of the current written in terms of these variables. The result (dropping both derivatives with respect to $\tilde t$ in
addition to $\tz$, for the reasons explained above) is
\be
J^{\tilde t}_{\widetilde{F1}} = \partial_\mu \left( \partial^\mu \xi^{\tilde t} + \partial^\mu H ( - \lambda_{\tz} +
\xi^{\tilde t} ) \right) \,.
\label{eq:Jtt}
\ee
One can check that this agrees with $J_t$ component of the current evaluated on the fundamental string by explicit calculation, setting $\lambda_{\tz} = \xi^z$ and $\xi^{\tilde t} = \lambda_t$, as expected from the T-duality. 

In the duality frame of the $\widetilde{\mathrm{F}1}$ it seems that we then have (up to our
normalisation)
\be
Q = \int \partial_\mu \partial^\mu H ( - \lambda_{\tz} + \xi^{\tilde t} ) \,.
\ee
These charges correspond to momentum in the dual direction, and the mass of this solution.
Interestingly, we have that $\xi^{\tilde t}$ appears with the opposite sign to $\xi^t$ in the original $J_{F1}^t$. This suggests that, having fixed our normalisation such that the F1 has positive mass, this timelike T-dual solution would appear to have negative mass if measured using this charge.

\subsection{Some remarks on magnetic charge}

In this subsection, we will make a few simple remarks about magnetic charges. These remarks will not
be especially novel, as they merely build on previous observations regarding the appearance of
fluxes in generalised geometry \cite{Ellwood:2006ya, Grana:2008yw} which have an obvious
generalisation in the frame or dynamical flux formulation
of double field theory \cite{
Siegel:1993xq, Hohm:2010xe,
Geissbuhler:2011mx,
Geissbuhler:2013uka }, but serve to
highlight the differences in defining such charges in double field theory compared to the electric-type
charges we have just been dealing with.

One of our motivations in considering the unfamiliar $\widetilde{F1}$  apparently generated by timelike T-duality was provided by our interest in the duality chain NS5-KKM-$5_2^2$. As we discussed, the NS5 brane is magnetically charged under $B_2$, with the KKM magnetically charged under
the Kaluza-Klein vector, and then the $5_2^2$ may be thought of as magnetically charged under the
bivector (as such it couples electrically to a mixed symmetry potential 
\cite{
Bergshoeff:2010xc,
Bergshoeff:2011zk, 
Bergshoeff:2011ee,
Bergshoeff:2011se,
Bergshoeff:2012ex
}).

As magnetic solutions, we would not expect to measure these using the charge we have found above.
Indeed, if one writes down for instance the explicit form of the NS5 solution
\be 
\begin{split}
ds^2 & = - dt^2 + d\vec{y}_5{}^2 + f d\vec{X}_4{}^2 \,,\\
 B_6 & = (1- f^{-1}) dt \wedge dy^1 \wedge\dots\wedge dy^5 \,,\\
e^{-2\phi} & = f^{-1} \,,
\end{split}
\label{eq:NS5}
\ee
where $f(|\vec{X}_4|)$ is a harmonic function of the transverse coordinates $\vec{X}_4$, then
it is straightforward to calculate that although the current $J^i$ has non-zero components, the
$J^t$ component is zero for constant $\xi^M$. Hence we cannot measure any (electric) charge using this current (though one can obtain the mass with appropriate treatment of boundary terms \cite{Park:2015bza, Naseer:2015fba}). However,
the NS5 has a magnetic charge under the $B$-field, $Q_m = \int H_3$. 

Thus, to measure this in a T-duality covariant manner in double field theory, one needs a
generalised analogue of the field strength of $B_2$. This is provided by a ``generalised flux''
\cite{Ellwood:2006ya, Geissbuhler:2013uka}. We may think of this as simply the generalised torsion of the Weitzenb\"ock connection:
\be
\tau_{MNP} = 3 \Omega_{[MNP]}  \,,
\ee
where we have lowered all indices using $\eta_{MN}$. 
When evaluated in terms of spacetime fields, this contains the flux of the two-form gauge field, so-called geometric flux and two non-geometric fluxes, $Q$- and $R$-flux \cite{Geissbuhler:2013uka} (see appendix \ref{appflux}). Thus, one can see that in double field theory we measure magnetic fluxes via the integral
\be
\int_{\Sigma_3} \tau_{MNP} dX^M \wedge dX^N \wedge dX^P \,.
\label{eq:inttau}
\ee
This integral is over some three-cycle in the doubled space: the choice of this three-cycle determines which particular type of flux one measures. Under T-duality, although the integrand is invariant, the three-cycle itself transforms. This can then be interpreted as the transformation of flux of one type to another under T-duality. From the doubled space point of view, there is no essential difference between the fluxes, it is only when we choose a physical section that we can label them as geometric or non-geometric, for instance.

Thus, by going to different T-duality frames (related by Buscher transformations, say), we obtain all the usual expressions for flux integrals (after integrating over the dual directions). Alternatively, by considering different choices of 3-cycles involving different numbers of dual directions we can in effect ``measure'' the presence of fluxes of different type.
This is just to say that we can successively obtain expressions of the form
\be
\int H_{ijk} dX^i dX^j dX^k \rightarrow
\int T_{i j}{}^k dX^i dX^j d\tilde X_k 
\rightarrow
\int Q_i{}^{jk} dX^i d\tilde X_j d\tilde X_k 
\rightarrow
\int R^{ijk} d \tilde X_i d\tilde X_j d\tilde X_k \,.
\ee
There are certain subtleties with the use of the generalised torsion in this way. In particular, the use of the vielbein means that one should be careful about the effect of local generalised Lorentz $O(D) \times O(D)$ transformations, under which this generalised torsion is not invariant. Such transformations are relevant in non-geometric backgrounds, as they allow one to switch between the description in terms of a $B$-field and that in terms of the bivector. To have a well-defined torsion integral, one should choose a vielbein which is globally defined only up to global $O(D,D)$ and (at most) constant $O(D) \times O(D)$ transformations.

For instance, consider the NS5 solution smeared in two transverse directions $X$ and $Y$. Going to
polar coordinates $(R,Z)$ in the other two transverse directions the harmonic function becomes a
logarithm in $R$ (divergent at some cut off $R_c$), and then (omitting the worldvolume
directions which play no role in this analysis) the solution can be written as 
\be
\begin{split} 
ds^2 & = f(R) \left( dR^2 + dX^2 + dY^2 + R^2 dZ^2 \right) \,,  \\
B_{XY} & = HZ \,,\\
e^{-2\phi} & = f^{-1}\,.
\end{split} 
\ee
where $X$, $Y$ are the compactified directions and $R,Z$ are cylindrical coordinates in the remaining transverse directions. Restricting our three-cycle $\Sigma_3$ to lie in the doubled space $(X,Y,Z,\tilde X, \tilde Y, \tilde Z)$, the only non-zero component of the generalised torsion is $\tau_{XYZ} = H$. Hence the charge integral tells us that the solution carries generalised flux, but only when we integrate over the $X,Y,Z$ directions - as our physical section is $(X,Y,Z)$ we interpret this as $B$-field flux. 

Now consider performing Buscher duality on the $X$, $Y$ directions. We obtain the non-geometric $5_2^2$ background \cite{deBoer:2012ma} with
\be
\begin{split}
ds^2 & = f K^{-1} \left( d\tilde X^2 + d\tilde Y^2 \right) + f ( dR^2 + R^2 dZ^2) \,, \\ 
B & = - HZ K^{-1} d \tilde X \wedge d \tilde Y \,, \\
e^{-2\phi} & = f^{-1} K  \,.
\end{split}
\ee
where $K = f^2 + H^2 Z^2$. This is defined only up to a global non-geometric T-duality for $Z
\rightarrow Z+2\pi$. The components of the generalised torsion are now $\tau_{Z \tilde X \tilde Y}
= - H K^{-1}$ and $\tau_{Z \tilde X}{}^{\tilde X} = \tau_{Z \tilde Y}{}^{\tilde Y} = -H^2 Z K^{-1}$.
This, though, is not well-behaved globally. 

To see why, 
note that the conventional choice of generalised vielbein in terms of the metric and $B$-field is not well-defined for the $5_2^2$ brane:
\be
E_{5_2^2}{}^\alpha{}_M = \begin{pmatrix} 
f^{1/2} K^{-1/2} \mathbb{I} & 0 \\
HZ K^{-1/2} f^{-1/2} \epsilon & f^{-1/2} K^{1/2} \mathbb{I} 
\end{pmatrix} \,.
\ee
One should instead use the dual frame, in which
\be
\begin{split}
ds^2 & = f^{-1} \left( d\tilde X^2 + d\tilde Y^2 \right) + f ( dR^2 + R^2 dZ^2) \,, \\ 
\beta^{\tilde X \tilde Y} &  = H Z \,, \\ 
e^{-2\phi} & = f \,. 
\end{split} 
\ee
In this frame, the generalised vielbein is 
\be
\widetilde E_{5_2^2}{}^\alpha{}_M = \begin{pmatrix}
f^{-1/2} \mathbb{I} & - f^{-1/2} HZ \epsilon \\
0 & f^{1/2} \mathbb{I}
\end{pmatrix} \,.
\ee
Whereas we have $E_{5_2^2}{}^\alpha{}_M (Z + 2 \pi) = P_M{}^N E_{5_2^2}{}^\beta{}_N (Z )
\Lambda_{\beta}{}^\alpha (Z)$ for $P_M{}^N \in O(D,D)$ a particular monodromy associated to the
background, and $\Lambda_{\beta}{}^\alpha (Z) \in O(D)
\times O(D)$, in the bivector frame we have $\widetilde E_{5_2^2}{}^\alpha{}_M (Z + 2 \pi) = P_M{}^N
\widetilde E_{5_2^2}{}^\alpha{}_N(Z) $. Hence for $\tau(Z) \equiv \tau_{ZMN} dX^M \wedge dX^N$ if we attempted to use the frame with the metric and
$B$-field we would have (as the coordinates themselves must be identified up to the global T-duality
$P_M{}^N$)
\be
\tau(Z+2\pi) = \tau(Z) + \partial_Z \Lambda_\beta{}^\alpha E_{\alpha N} E^\beta{}_M dX^M \wedge dX^N
\,,
\ee
while in the non-geometric frame we have $\widetilde \tau(Z+2\pi) = \widetilde\tau(Z)$. So the
generalised torsion is only well-defined globally in this case, with the only relevant non-zero component being
\be
\widetilde\tau_{Z}{}^{\tilde X \tilde Y} = H \,.
\ee
Thus we only obtain a non-zero charge when integrating over a single physical direction $Z$ and over two dual directions ($X$ and $Y$) - this is interpreted as non-geometric $Q$-flux. 

\section{Additional terms in the action}
\label{sec:5}

In this final section, we will consider the effects of adding additional terms to the action of double
field theory. 

\subsection{Total derivatives}

The simplest possibility is to consider adding a total derivative to the action
\be
S_\Delta = \int \partial_M \left( e^{-2d} K^M \right) \,,
\ee
where $K^M$ is assumed to be a vector under generalised diffeomorphisms. It is straightforward to
show that this leads to an additional contribution to the current
\be
\begin{split} 
J_\Delta^M & = 
\nabla_N \left( 2 e^{-2d} K^{[M} \xi^{N]} \right) 
+ \eta^{MN} \eta_{PQ} K^Q \nabla_N \xi^P \\ 
& = 
\partial_N \left( 2 e^{-2d} K^{[M} \xi^{N]} \right) 
+ \eta^{MN} \eta_{PQ} K^Q \partial_N \xi^P \,.
\end{split}
\ee

\subsection{Scherk-Schwarz term} 
\label{se:SS} 

An alternative to the explicit solution $\tilde \partial^i =0$ of the section condition is to use a Scherk-Schwarz reduction
\cite{
Aldazabal:2011nj, 
Geissbuhler:2011mx,
Grana:2012rr}. The coordinates are split into doubled ``internal'' and doubled ``external''
sets, denoted now $\mathbb{X}$ and $\mathbb{Y}$, and the generalised vielbein and dilaton are
decomposed via the reduction ansatz $E^\alpha{}_M ( \mathbb{X} , \mathbb{Y} ) = U_M{}^A(\mathbb{Y})
\hat{E}^\alpha{}_A(\mathbb{X})$ and $d ( \mathbb{X} , \mathbb{Y} ) = \hat{d} ( \mathbb{X} ) +
\lambda(\mathbb{Y})$. The physical fields and gauge parameters of the reduced theory - which is
termed a gauged double field theory - will depend
only on $\mathbb{X}$ (and are denoted with hats). All dependence on the internal coordinates
$\mathbb{Y}$ through the twist matrices $U_M{}^A$ and dilaton twist $\lambda$ is absorbed into
generalised fluxes (providing the gaugings of the theory), which are taken to be constant. If we make use of the Weitzenb\"ock connection,
these fluxes emerge naturally from the torsion and the covariant derivative of the generalised
dilaton:\cite{Berman:2013uda}
\be
\tau_{MNP} ( \mathbb{X} , \mathbb{Y} ) = U_M{}^A U_N{}^B U_N{}^C \left( \hat{\tau}_{ABC} (
\mathbb{X} ) + f_{ABC}\right) \quad , \quad
D_M d = U_M{}^A \left( \hat{D}_A \hat{d} ( \mathbb{X} ) + f_A \right) \,,
\ee
where 
\be
f_{ABC} = 3 U^M{}_{[A} U^N{}_B \partial_{|M} U_{N| C]} \quad , \quad
f_A = \partial_M U^M{}_A - 2 U^M{}_A \partial_M \lambda \,.
\ee
After imposing this ansatz - plus some other consistency conditions \cite{Grana:2012rr} -  one finds that the section condition on the full $(\mathbb{X},
\mathbb{Y})$ space can be replaced by the section condition for the external coordinates and set of
Jacobi identity-like constraints involving the fluxes, which can be shown to be less restrictive
than the original section condition. 

Given our conserved current $J^M$ - which is a generalised vector of weight one - it is straightforward to check that the conservation law
$\partial_M J^M = 0$ gives
\be
\partial_M J^M = e^{-2\lambda} \left( \partial_A \hat{J}^A + f_A \hat{J}^A \right) = 0 \,.
\ee
It turns out \cite{Grana:2012rr} that $f_A$ must be taken to vanish for consistency of the
Scherk-Schwarz (although this can be relaxed in more complicated versions of the reduction ansatz
\cite{Geissbuhler:2011mx}). Hence given any conserved current $J^M$ we will obtain also a conserved
current $\hat{J}^A$ in the resulting
gauged double field theory. If the current can be written entirely in terms of the Weitzenb\"ock
connection and its torsion (as in \eqref{eq:JWeitz}), then it will take essentially the same form in
the gauged double field theory, with all indices
$M,N,\dots$ replaced by indices in the gauged double field theory, $A,B, \dots$ and all tensors
appearing as hatted quantities, and the fluxes appearing through the replacement $\tau_{MNP} \rightarrow \hat{\tau}_{ABC} +
f_{ABC}$. 

In order to derive the full gauged double field theory via a reduction of this type, however, one
needs to make a small modification to the original double field theory action \cite{Grana:2012rr}.
One adds to the action a term
\be
S_\eta \equiv \frac{1}{2} \int e^{-2d} \eta^{MN} \partial_M E^\alpha{}_P \partial_N E_{\alpha}{}_Q \eta^{PQ} \,.
\ee
We call this the Scherk-Schwarz term. In the ordinary formulation of double field theory, it
vanishes identically by the section condition. However, it makes necessary contributions to the
reduced action when one imposes instead the Scherk-Schwarz ansatz. Therefore, to consider the true
conserved current in a Scherk-Schwarz setting we must include this term. 

If we proceed to vary this directly, we obtain
\be 
\begin{split}
\delta S_\eta & = - \int e^{-2d} \eta^{MN} \eta^{PQ} (\partial_M - 2 \partial_M d) \partial_N E_{\alpha Q} \delta E^\alpha{}_P \\ 
& + \int \partial_M \left( e^{-2d} \eta^{MN} \eta^{PQ} \delta E^\alpha{}_P \partial_N E_{\alpha Q} \right) \,.
\end{split}
\ee
(The only effect of the variation of the generalised dilaton simply shifts the Lagrangian appearing in the Bianchi
identity.) From this, the current picks up a contribution
\be
\begin{split} 
e^{-2d} \eta^{MN} \Omega_{NKL} \left( D_P \xi^K \gM^{LP} + \eta^{PK} S^L_Q D_P \xi^Q 
- \xi^S \tau_{SP}{}^K \gM^{LQ} \right) \\ 
+ 2 e^{-2d} \eta^{M[N} \xi^{P]} \eta^{KL} E^\alpha{}_N ( \partial_K -2\partial_K d) \partial_L
E_{\alpha P} \,.
\end{split}
\ee
This is not especially pleasant and does not give a covariant modification of the current. 

It is more natural to study the Scherk-Schwarz term using the flux or Weitzenb\"ock formulation 
\cite{Geissbuhler:2011mx, Berman:2013uda}. Otherwise, one has to view it as being added to the
original DFT action by hand, as originally done in \cite{Grana:2012rr} in order to obtain consistent Scherk-Schwarz compactifications. 

In this approach, the DFT action including the Scherk-Schwarz term can be written as $S = \int dx d\tilde x e^{-2d} L$ with
\be
\begin{split}
L & = 
-\frac{1}{12} \gM_{MQ} \gM^{NR} \gM^{PS} \tau_{NP}{}^M \tau_{RS}{}^Q
-\frac{1}{4} \gM^{NQ} \tau_{NP}{}^M \tau_{QM}{}^P
\\ & 
- 4 \gM^{MN} D_M d D_N d
+ 4 \gM^{MN} D_M D_N d
\,.
\end{split}
\label{eq:WeitzL}
\ee
This is the sole combination quadratic in the generalised torsion and covariant derivative of the generalised dilaton which is invariant under the local $O(D)\times O(D)$ symmetry, up to section condition. If one varies this form of the action under generalised diffeomorphisms as before one finds at the end of the calculation the identity
\be
\begin{split}
0   &  = \int dx d\tilde xe^{-2d} \xi^M 
\Bigg(
Z_{MNP}{}^Q
\left(
\frac{1}{6} \gM_{QS} \gM^{KN} \gM^{LP} \tau_{KL}{}^S + \frac{1}{2} \gM^{SN} \tau_{SQ}{}^P 
\right) 
 \\ 
  & + 
\int dx d\tilde x \partial_M J^M \,.
\end{split}
\ee
where
\be
Z_{MNP}{}^Q \equiv
3 D_{[M} \tau_{NP]}{}^Q - \eta_{MT} \eta^{QR} D_R \tau_{NP}{}^T 
-3 \tau_{[MN}{}^R \tau_{P]R}{}^Q = 0  
\ee
is the Bianchi identity for the generalised torsion of the Weitzenb\"ock connection \cite{Berman:2013uda}, and the current that one obtains can be written
\be
J^M = D_{N} J^{MN} + \frac{1}{2} \tau_{KL}{}^M J^{KL} \,,
\ee
with
\be
J^{MN} = e^{-2d} \xi^P \left( 
\left( \eta^{MQ} \eta^{NR} - \gM^{MQ} \gM^{NR} \right) \gM_{PS} \tau_{QR}{}^S 
+ \gM^{MS} \tau_{PS}{}^N - \gM^{NS} \tau_{PS}{}^M  
\right)  \,.
\ee
This current differs somewhat from that obtained from the original DFT action. One of the main differences is that there are no derivatives of $\xi^P$ in $J^{MN}$. 

In the original formalism, working through the variation one encounters terms of the schematic form $\partial \delta_\xi \gM$ and $\partial \delta_\xi d$ appearing in $J^M$ (see equation \eqref{eq:VaryResult0}). As generalised diffeomorphisms involve derivatives of $\xi^P$, one therefore finds terms in $J^M$ involving two derivatives of $\xi^P$. 

In the formalism we are considering in this subsection, one finds instead after integration by parts
that the relevant term apparently containing the most derivatives is $\delta_\xi D_P d$. However, $D_P d$ is tensorial. Hence we only obtain single derivatives of $\xi^P$ in $J^M$ when we substitute in the expressions for generalised diffeomorphisms. 

Ultimately, this discrepancy is a result of a subtlety involving two derivative terms in the action.
In the formulation using solely the generalised metric, we have a term $\partial_M \partial_N
\gM^{MN}$. In terms of the vielbein though, this gives rise to the following possibilities
\be
\begin{split} 
\int  e^{-2d} E_\alpha{}^M \partial_M \partial_N \delta E^{\alpha N} & = 
\int  \partial_M \left( e^{-2d} E_\alpha{}^M \partial_N \delta E^{\alpha N} \right) 
- \int \partial_M \left( e^{-2d} E_\alpha{}^M \right) \partial_N \delta E^{\alpha N} \\ 
& = 
\int  \partial_M \left( e^{-2d} E_\alpha{}^N \partial_N \delta E^{\alpha M} \right) 
- \int \partial_M \left( e^{-2d} E_\alpha{}^N \right) \partial_N \delta E^{\alpha M}\,.
\end{split} 
\ee 
Here we have a choice of which partial derivative in the following two-derivative term we use to
partially integrate. 
By picking one or the other, although the total variation is the same, one alters which terms appear
in the total derivative and which do not. Effectively, when working with the Lagrangian
\eqref{eq:WeitzL} one ends up being forced into making a particular choice of derivative. A consequence of this is to ensure that all parts of the variation are written in terms of the covariant quantities $D_M d$ and $\tau_{MN}{}^P$ and hence that the Bianchi identities involve these. 

In fact this same subtlety arises in general relativity, when comparing the current obtained from
the usual formulation with that obtained by varying the teleparallel action which uses the torsion
of the spacetime Weitzenb\"ock connection. 

Note also that in the Bianchi identities, one has appearing first order derivatives of geometric quantities which are themselves first order in derivatives. In the original Bianchi identities involving the generalised torsion-free connection, one has appearing first order derivatives of geometric quantities which are second order in derivatives. Thus, it may make sense from this point of view for there to have occurred some rearrangement of the placement of the terms which are (overall) third-order in derivatives in our variation. 

\subsection{Coupling to RR fields}

\subsubsection{The RR action}

The incorporation of RR fields into double field theory was originally achieved in the papers \cite{Hohm:2011zr, Hohm:2011dv}, and adapted to the flux formulation in \cite{Geissbuhler:2011mx} (and see also \cite{Jeon:2012kd}).  Here, we will follow the presentation of \cite{Geissbuhler:2011mx,Geissbuhler:2013uka} (and refer the reader to \cite{Hohm:2011dv} for more general facts about $O(D,D)$ spinors). 

We introduce $O(D,D)$ Gamma matrices carrying \emph{flat} indices
\be
\{ \Gamma^\alpha, \Gamma^\beta \} = 2 \eta^{\alpha \beta} \,. 
\ee
One can use the generalised vielbein to obtain curved gamma matrices, $\Gamma^M \equiv E_\alpha{}^M
\Gamma^\alpha$. As we require $E_\alpha{}^M E_\beta{}^N \eta^{\alpha \beta} = \eta^{MN}$, these also
obey the above defining relation, but for curved indices. 

A practical realisation of these gamma matrices is to introduce $D$ pairs of fermionic creation and annihilation operators $\psi_\mu,
\psi^\nu$ such that $(\psi^\mu)^\dagger = \psi_\mu$ and
\be
\{ \psi^\mu, \psi^\nu \} = \{ \psi_\mu , \psi_\nu \} = 0 \quad \{ \psi_\mu, \psi^\nu \} =
\delta^\mu_\nu \,.
\ee
Then we have $\Gamma^\alpha = ( \sqrt{2} \psi^\mu , \sqrt{2} \psi_\mu )$. We can construct a general
spinor by introducing a vacuum state $|0\rangle$ annihilated by the $\psi_\mu$ and defining 
\be
 \lambda = \sum_{p=0}^{D} \frac{1}{p!} \lambda_{\mu_1  \dots  \mu_p} \psi^{\mu_1}
\dots \psi^{\mu_p} |0 \rangle \,.
\ee
One can immediately see then that an $O(D,D)$ spinor should correspond to a set of $p$-forms in spacetime. 

The charge conjugation matrix is
\be
\Cc = ( \Gamma^0 \pm \Gamma_0 ) \dots ( \Gamma_{D-1} \pm \Gamma^{D-1} )\,,
\ee
where we use plus signs for $D$ odd and the minus signs for $D$ even.
Letting
\be
\Gamma^{\alpha_1 \dots \alpha_n} = \Gamma^{[\alpha_1} \dots \Gamma^{\alpha_n]} \,,
\ee
we have
\be
\Cc^T = (-1)^{D(D-1)/2} \Cc \quad
( \Cc \Gamma^{\alpha_1 \dots \alpha_n})^T = (-1)^{(D-n)(D-n+1)/2} \Cc \Gamma^{\alpha_1 \dots
\alpha_n} \,.
\ee
We can now turn to the inclusion of the RR fields of supergravity. We encode these in the spinor
\be
C = \sum_{p=0}^D e^\phi C_{i_1 \dots i_p} e^{i_1}_{\mu_1} \dots e^{i_p}_{\mu_p} \psi^{\mu_1} \dots
\psi^{\mu_p} |0\rangle \,,
\ee
where $i$ is a curved spacetime index and $\phi$ is the usual dilaton. 
Next, define a Dirac operator
\be
\Dop = \frac{1}{\sqrt{2}} \Gamma^M \partial_M - \frac{1}{\sqrt{2}} \Gamma^M D_M d +
\frac{1}{12\sqrt{2}} \Gamma^{MNP} \tau_{MNP} \,,
\ee
using the Weitzenb\"ock connection $\Omega_{MN}{}^P$ and its torsion. Note we lower the index on the
torsion using $\eta$. This operator obeys 
\be
\Dop \Dop = 0 \,,
\ee
by the section condition. 

The field strengths are then encoded by
\be
G = \Dop C \,.
\ee
By reducing to components one can check that if $C$ represents the polyform $e^{\phi} \sum C_{(p)}$ then with
these definitions $G$ gives the polyform $e^{\phi} (d + H_3 \wedge) \sum C_{(p)}$. 

The (pseudo)-action is then \cite{Geissbuhler:2011mx}
\be
S_{RR} = - \frac{1}{4} \int dx d\tilde x e^{-2d} \bar{G} \Psi_+ G \,.
\ee
Here
\be
\Psi_+ = ( \Gamma^0- \Gamma_0 ) ( \Gamma^1 + \Gamma_1 ) \dots ( \Gamma^{D-1} + \Gamma_{D-1} ) \,,
\ee
is the spin representative of the flat generalised metric, and obeys
\be
\Psi_+^{-1} = \Psi_+^T = (-1) (-1)^{D(D-1)/2} \Psi_+ \,,
\ee
and
\be
\Psi_+ \Cc = - \Cc \Psi_+ \,.
\ee
Finally, the action must be supplemented with a self-duality condition which is to be imposed after
varying. This condition is just
\be
G = \Psi_+ G \,.
\ee
In order to derive the contribution of the coupling to the RR fields to the current associated to
generalised diffeomorphisms, we now proceed as before by varying the action. 
We have
\be
\begin{split} 
\delta S_{RR} & = \frac{1}{2 \sqrt{2}} \int dxd\tilde x \partial_M 
\Bigg( e^{-2d} 
\delta d  \bar{C} \Gamma^M \Psi_+ G -   e^{-2d} \delta \bar{C} \Gamma^M \Psi_+ G 
\\ & \qquad\qquad
+  e^{-2d} E_\alpha{}^P  \delta E^\alpha{}_N
\left( \frac{1}{2} \delta_P^M \bar{C} \Gamma^N \Psi_+ G + \frac{1}{4} \bar{C} \Gamma^{MN}{}_P
\Psi_+ G \right) 
\Bigg)
\\ & 
 - \frac{1}{2} \int dx d\tilde x e^{-2d} 
\Bigg( 
\delta d \bar{C} \Dop \Psi_+ G  - \delta \bar{C} \Dop \Psi_+ G 
\\ 
& -  \delta E_\alpha{}^P  \eta^{\alpha \beta} E_\beta{}^N  \left(
\frac{1}{4} \eta_{QN} \eta_{PR} \bar{C} \Gamma^{QR} \Dop \Psi_+ G 
+ \frac{1}{4} \eta_{QN} \eta_{PR} \bar{G} \Gamma^{QR} \Psi_+ G 
\right) \Bigg) \,.
\end{split}
\ee
Using the self-duality relation we see that there is no contribution to the generalised dilaton
equation of motion, while the vielbein equation of motion picks up an additional term (note that
$\delta E_{\alpha}{}^P \eta^{\alpha \beta} E_\beta{}^N$ must be antisymmetric in $PN$ so that the
coset condition is preserved).

\subsubsection{Variation under generalised diffeomorphisms}

We have
\be
\delta_\xi d = \xi^M \partial_M d - \frac{1}{2} \partial_M \xi^M \,,
\ee
\be
\delta_\xi E_\alpha{}^P  = \xi^M \partial_M E_\alpha{}^P - E_\alpha{}^M \partial_M \xi^P + \eta^{PM}
\eta_{QR} \partial_M \xi^Q E_\alpha{}^R  \,,
\ee
while in this formulaton the RR spinor is a scalar under generalised diffeomorphisms
\cite{Geissbuhler:2011mx, Geissbuhler:2013uka}
\be
\delta_\xi C = \xi^M \partial_M C \,.
\ee
Then starting with the above variations and dropping the terms involving $\Dop \Psi_+ G$ (which
vanish on applying the self-duality condition) we find
\be
\begin{split}
\delta_\xi S_{RR} & = - \frac{1}{4} \int \partial_M \left( \xi^M e^{-2d} \bar{G} \Psi_+ G \right) \\
 & = \frac{1}{2\sqrt{2}} \int \partial_M 
\left( \xi^M \delta_\xi d \bar{C} \Gamma^M \Psi_+ G - \delta_\xi \bar{C} \Gamma^M \Psi_+ G \right.
\\ & \left.
+ E_\alpha{}^P \delta_\xi E^\alpha{}_M \left( \frac{1}{2} \delta_P^M \bar{C} \Gamma^N \Psi_+ G +
\frac{1}{4} \bar{C} \Gamma^{MN}{}_P \Psi_+ G
\right) \right) \\
 & + \frac{1}{4} \int \partial_M \left( e^{-2d} \xi^N \eta_{QN} \bar{G} \Gamma^{QM} \Psi_+ G \right) 
\\
& + 
\int dx d\tilde x e^{-2d} \xi^N \sqrt{2} \left( - \frac{1}{4} \bar{G} \Gamma^Q \Dop \Psi_+ G
\eta_{QN} + \frac{1}{4}\bar{\Dop G} \Gamma^Q \Psi_+ G \eta_{QN} \right) 
\end{split} 
\ee
The last line vanishes by the Bianchi identity $\Dop G = 0$ (and again on using the self-duality
condition). From the rest we obtain the conserved current to be
\be
\begin{split}
J_{RR}^M & = \frac{1}{2 \sqrt{2} } e^{-2d} \xi^N ( \nabla_N \bar{C} - \nabla_N d \bar{C} ) \Gamma^M
\Psi_+ G 
- \frac{1}{4 \sqrt{2}} \nabla_N \xi^N \bar{C} \Gamma^M \Psi_+ G 
\\
& + \frac{1}{8\sqrt{2}}
\left( \eta_{PQ} \nabla_N \xi^P - \eta_{PN} \nabla_Q \xi^P + \tau_{PNQ} \xi^P\right)  \bar{C} \Gamma^N
\Gamma^Q \Gamma^M \Psi_+ G
\\
&+ \frac{1}{4} \xi^M \bar{G} \Psi_+ G 
+ \frac{1}{4} \xi^N \eta_{QN} \bar{G} \Gamma^{QM} \Psi_+ G 
\end{split}
\ee
A certain amount of manipulation (including using the self-duality relation and the Bianchi
identity to drop certain terms)
leads us to the more compact expression
\be
\begin{split} 
J_{RR}^M & = 
D_N J_{RR}^{MN}
+ \frac{1}{2} \tau_{KL}{}^M J_{RR}^{KL} + \varphi^M\\
& = \partial_N J_{RR}^{MN} + \frac{1}{2} \eta^{MP} \eta_{LQ} \Omega_{PK}{}^Q J_{RR}^{KL} 
+ \varphi^M 
\,,
\end{split}
\ee
with
\be
J_{RR}^{MN} = \frac{1}{4 \sqrt{2}} e^{-2d} \xi^P \eta_{PQ} \bar{C} \Gamma^{QMN} \Psi_+ G 
+ \frac{1}{2\sqrt{2}} e^{-2d} \xi^{[M} \bar{C} \Gamma^{N]} \Psi_+ G \,,
\ee
and the terms which vanish by the section condition in $\partial_M J_{RR}^M$ are
\be
\varphi^M = \frac{1}{2 \sqrt{2}}e^{-2d}  \eta^{MN} \eta_{PQ} \xi^P \bar{C} \Gamma^Q \partial_N \Psi_+ G 
- \frac{1}{4\sqrt{2}} \eta^{MN}e^{-2d} \eta_{PQ}  \partial_M \left( \xi^P \bar{C} \Gamma^Q \Psi_+ G \right)
  \,.
\ee

\subsubsection{Variation under $C$-field gauge transformations}

We may also derive here the current that is associated to the gauge transformations of the RR
sector. These are written in spinor language as
\be
\delta C = \Dop \lambda \,,
\ee
and the field strenth $G = \Dop C$ is then invariant as $\Dop \Dop = 0$ by section condition. Under
these transformations, we have
\be
\begin{split} 
\delta_\lambda S_{RR} & = - \frac{1}{2\sqrt{2}} \int \partial_M \left(
 e^{-2d} \overline{\Dop \lambda} \Gamma^M\Psi_+ G \right) 
+ \frac{1}{2} \int dxd\tilde x e^{-2d} \overline{\Dop \lambda}  \,\Dop \Psi_+ G \\
\\ & = \frac{1}{2\sqrt{2}} \int dxd\tilde x \partial_M \Big(
 e^{-2d} \overline{\lambda} \Gamma^M \Dop \Psi_+ G 
-e^{-2d} \overline{\Dop \lambda} \Gamma^M\Psi_+ G 
\Big) 
\\ & 
- \frac{1}{2\sqrt{2}} \int dx d\tilde x e^{-2d} \overline{\lambda} \, \Dop \Dop \Psi_+ G \,.
\end{split} 
\ee
We thus see that invariance under $C$-field gauge transformations implies the identity $\Dop \Dop =
0$, which is a generalised version of the usual exterior derivative identity $d^2 = 0$. We can also then simplify the
conserved current, which takes a by-now familiar form:
\be
\begin{split}
J_\lambda^M 
& =
D_N J_\lambda^{MN} + \frac{1}{2} \tau_{KL}{}^M J_\lambda^{MN} \\ & 
- \eta^{MN} \frac{1}{2\sqrt{2}} e^{-2d} \partial_N \overline{\lambda} \Psi_+ G 
+ \eta^{MN} \frac{1}{2\sqrt{2}} e^{-2d} \overline{\lambda} \partial_N \Psi_+ G \,,
\end{split} 
\ee
with
\be
J_\lambda^{MN} = \frac{1}{2\sqrt{2}} e^{-2d} \overline{\lambda} \Gamma^{MN} \Psi_+ G \,.
\ee

\section{Conclusions}
\label{sec:6}

In this work, our main focus was to calculate the conserved currents associated to generalised
diffeomorphisms in the NSNS sector of double field theory. We obtained the off-shell conserved
current, from which one can define conserved charges unifying Komar-type charges with the electric
charges associated to the $B$-field gauge invariance in a T-duality covariant manner. This allowed us
to formalise the double field theory interpretation of the fundamental string solution as carrying
momentum in a dual direction \cite{Berkeley:2014nza}. We also extended our results to the case where
the action was supplemented with the Scherk-Schwarz term, and to the inclusion of the RR sector.

A very natural next step is then to continue calculating this current for the other extensions of
double field theory, such as the heterotic double field theory \cite{Hohm:2011ex} and the
supersymmetric extensions, in which the full type II supergravities are unified \cite{Jeon:2012hp}.
There it would be interesting to consider the algebra of charges and supercharges. 
In addition, one
could look at the recent formulation of double field on group manifolds
\cite{
Blumenhagen:2014gva,
Blumenhagen:2015zma
}. 

The other obvious - and perhaps more interesting - extension is the usual generalisation from T- to
U-duality. It should not be difficult to calculate the current associated to the generalised
diffeomorphisms which appear in an exceptional extended geometry 
\cite{
Berman:2010is, 
Aldazabal:2013mya,
Hohm:2013vpa, 
Hohm:2013uia, 
Hohm:2014fxa, Godazgar:2014nqa,
Berman:2014jsa,Blair:2014zba}. One reason this is interesting is that one will have
duality transformations which map electric and magnetic solutions into each other (for instance, the
M2 and M5). Indeed, it has been shown that for the $E_7$ all the familiar half-BPS solutions appear as a single self-dual solution of exceptional field theory \cite{Berman:2014hna}. This is unlike the case in T-duality, where we have argued that one needed different
definitions of charge for electrical solutions like the F1 and magnetic solutions like the NS5. It
is possible already to study the generalised fluxes of some of these exceptional field theories
\cite{Blair:2014zba}, and so it would be interesting to see what is contained additionally in the
electrical current. 

We should also return to the conserved currents of the RR sector in more detail. We presented only their
derivation in section \ref{sec:5}. In particular, one would expect after reducing to spacetime that the
combination of the NSNS and RR sector charges will reproduce more complex definitions of electric
charges relevant to situations with multiple sources. The relationship of this to Page charge \cite{Page:1984qv, Marolf:2000cb}, should be investigated, as this is the charge relevant when considering realistic examples of exotic brane configurations and their monodromies \cite{deBoer:2012ma, Kikuchi:2012za, Okada:2014wma}. Double field theory should provide a natural setting in which to study such non-geometric branes, in which case our expressions for the charge should come into play in an interesting way.

\section*{Acknowledgements}

The author would like to acknowledge useful discussions with Emanuel Malek in particular, and also
David Berman, Malcolm Perry and Daniel Thompson, and would also like to thank Jeong-Hyuck Park for
drawing his attention to the related work \cite{Park:2015bza}. The author is grateful for the support of St John's College,
Cambridge. 

\appendix 

\section{The $O(D,D)$ geometry of double field theory}

In this appendix, we collect useful results on specific connections in double field theory, many of which are needed in deriving the forms of the conserved currents given in the paper. 
We will freely raise and lower using $\eta$, except where noted. 

\subsection{The semi-determined/semi-covariant connection} 

\subsubsection{The connection and its properties} 

The connection which most closely resembles the Levi-Civita connection is that explored by
\cite{Jeon:2010rw, Jeon:2011cn, Hohm:2011si}. We will more closely follow the notation of the paper
\cite{Hohm:2011si}. Firstly, we
define the projectors
\be
P_M^N = \frac{1}{2} ( \delta - S)_M^N \quad , \quad  
\bar{P}_M^N = \frac{1}{2} ( \delta + S)_M^N \,,
\ee
and note that we may occasionally use the notation
\be
A_{ \underline{M}} \equiv P_M^N A_N \quad , \quad 
A_{\overline M} \equiv \bar{P}_M^N A_N \,.
\ee
The connection by definition annihilates the generalised metric, the $O(D,D)$ structure and the
generalised dilaton. The latter two conditions imply
\be
\Gamma_{MNP} = - \Gamma_{MNP} 
\ee
and
\be
\Gamma_{NM}{}^N = - 2 \partial_M d \,.
\ee
The connection is also free of generalised torsion, which implies
\be
\Gamma_{MNP} 
+\Gamma_{NPM} 
+\Gamma_{PMN}
= 0 \,. 
\ee
The combination of the latter two constraints gives
\be
\eta^{MN} \Gamma_{MNP} = 2 \partial_P d 
\ee
The explicit connection constructed has the form
\be
\begin{split}
\Gamma_{MNK} & = - 2 P_{[N|}^Q \partial_M P_{Q|K]} 
- 2 \bar{P}_{[N}^P \bar{P}_{K]}^Q \partial_P P_{QM} 
+ 2 {P}_{[N}^P {P}_{K]}^Q \partial_P P_{QM} \\
& + \frac{4}{D-1} \left( 
P_{M[N} P_{K]}^Q 
+ \bar P_{M[N} \bar P_{K]}^Q  \right) 
\left( \partial_Q d + P_{[P}^L \partial^P P_{|L|Q]} \right) \\ 
& + \widetilde \Gamma_{MNK} \,.
\end{split} 
\ee
The undetermined components at the end vanish when projected by both $P$ and $\bar{P}$. One may also
write
\be
\begin{split}
\Gamma_{MNK} & = 
\frac{1}{2} \gM_{KQ} \partial_M S^Q_N
+ \frac{1}{2} \left( \delta_{[N}^P S_{K]}^Q + S_{[N}^P \delta_{K]}^Q \right) \partial_P \gM_{QM}
\\ & 
+ \frac{2}{D-1} \left( \eta_{M[N} \delta_{K]}^Q + \gM_{M[N} S_{K]}^Q \right) \left(
\partial_Q d + \frac{1}{4} \gM^{PM} \partial_M \gM^{PQ} 
\right) \\ 
& + \widetilde \Gamma_{MNK} \,.
\end{split}
\ee
It is straightforward to check the following contractions:
\be
\eta^{MN} \Gamma_{MNP} = 2 \partial_P d \quad , \quad \gM^{MN} \Gamma_{MNP} = 2 S_P^Q \partial_Q d -
\partial_Q S^Q_P \,,
\ee
\be
P^{MN} \Gamma_{MNP} = 2 P_P^Q \partial_Q d - \partial_Q P^Q_P \quad , \quad
\bar{P}^{MN} \Gamma_{MNP} = 2 \bar{P}_P^Q \partial_Q d - \partial_Q \bar{P}_P^Q \,,
\ee
\be
\Gamma_{MNP} P_R^N \bar{P}_S^P = - \bar{P}_R^Q \partial_M P_{QS} \,,
\ee
\be
P_R^M \bar P_S^N \Gamma_{MNP} = - P_R^Q P_{PN} \partial_Q\bar{P}_S^N 
+ \bar{P}_P^Q \bar{P}_{SM} \partial_Q P_R^M
- \bar{P}_S^Q \bar{P}_{PM} \partial_Q P_R^M\,,
\ee
\be
\bar P_R^M P_S^N \Gamma_{MNP} = - \bar P_R^Q \bar P_{PN} \partial_Q P_S^N 
+ {P}_P^Q {P}_{SM} \partial_Q \bar P_R^M
- {P}_S^Q {P}_{PM} \partial_Q \bar P_R^M\,.
\ee

\subsubsection{Generalised Riemann tensor and Ricci identity} 

Now, define the conventional Riemann tensor
\be
R_{MNK}{}^L = \partial_M \Gamma_{NK}{}^L -
\partial_N \Gamma_{MK}{}^L
+ \Gamma_{MQ}{}^L \Gamma_{NK}{}^Q 
- \Gamma_{NQ}{}^L \Gamma_{MK}{}^Q \,,
\ee
such that 
\be
[ \nabla_M , \nabla_N ] A^P =  R_{MNL}{}^P A^L 
- \tau_{MN}{}^L \nabla_L A^P 
+ \eta^{LR} \eta_{NS} \Gamma_{RM}{}^S \nabla_L A^P \,,
\ee
where $\tau_{MN}{}^L$ is the generalised torsion. The generalised Riemann tensor is then
\be
\mathcal{R}_{MNK}{}^L = R_{MNK}{}^L + \eta^{LP} \eta_{NQ} R_{KPM}{}^Q 
+ \eta^{QR} \eta_{NP} \Gamma_{QM}{}^P \Gamma_{RK}{}^L \,.
\ee
One has the following useful identities (assuming the annihilation of the $O(D,D)$ structure and
generalised metric):
\be
R_{MNKL} = - R_{NMKL} = - R_{MNLK} \,,
\ee
\be
R_{MN \underline{K} \overline{L} } = 0 \,,
\ee
\be
\mathcal{R}_{MNKL} = \mathcal{R}_{KLMN} \,,
\ee
\be
\mathcal{R}_{MNKL} = - \mathcal{R}_{NMKL} = - \mathcal{R}_{MNLK} \,,
\ee
\be
\mathcal{R}_{\underline{M} \underline{N} \overline{K} \overline{L}} = 0 
\quad , \quad 
\mathcal{R}_{\underline{M} \overline{N} \underline{K} \overline{L}} = 0 \,.
\ee
One also has a Bianchi identity assuming the generalised torsion vanishes:
\be
\mathcal{R}_{[MNK]L} = 0 \,.
\ee
To obtain a Ricci-like identity for a generalised torsion free connection, we can rewrite 
\be
\begin{split} 
[ \nabla_M , \nabla_N ] A^P & =  \mathcal{R}_{MNL}{}^P A^L 
- \eta^{PK} \eta_{NQ} R_{LKM}{}^Q A^L 
+ \eta^{LR} \eta_{NS} \Gamma_{RM}{}^S \partial_L A^P \\ & 
=  \mathcal{R}_{MNL}{}^P A^L 
- \eta^{PK} R_{LKMN}{}^Q A^L 
+ \eta^{LR} \Gamma_{RMN} \partial_L A^P 
\end{split} 
\ee
and project
\be
\begin{split} 
P_M{}^K \bar{P}_N{}^L
[ \nabla_K , \nabla_L ] A^P &  =  \mathcal{R}_{\underline{M} \overline{N} L}{}^P A^L 
- \eta^{PK} R_{LK \underline{M} \overline{N} } A^L 
+ \eta^{LR} \Gamma_{R \underline{M} \overline{N} } \partial_L A^P \\
& =  \mathcal{R}_{\underline{M} \overline{N} L}{}^P A^L \,,
\end{split} 
\ee
using the identity $R_{LK \underline{M} \overline{N} } = 0$ and the fact that
\be
\eta^{LR} \Gamma_{R \underline{M} \overline{N} } \partial_L A^P 
= - \eta^{LR} \bar P_M^Q \partial_R P_{QN} \partial_L A^P \,,
\ee
which vanishes by the section condition. 

\subsubsection{Generalised Ricci tensor}
\label{app:Rbi}

The generalised Ricci tensor is defined as
\be
\mathcal{R}_{MN} = \left( P_M^K \bar{P}_N^L + P_N^K \bar{P}_M^L \right) P_P^Q \mathcal{R}_{QKL}{}^P 
= \left( P_M^K \bar{P}_N^L + P_N^K \bar{P}_M^L \right) \bar P_P^Q \mathcal{R}_{QKL}{}^P 
\ee
One then has a contracted Ricci identity
\be
\left( P_M^K \bar{P}_N^L + \bar{P}_M^K P_N^L \right) [ \nabla_K , \nabla_L ] A^N = - R_{MN} A^N \,.
\label{eq:contractedRicci}
\ee
The generalised Ricci tensor obeys the following Bianchi identities:
\be
\nabla_{\underline{P}} \mathcal{R} - 4 \nabla^{\overline{M}} \mathcal{R}_{\underline{P} \overline{M}} =
0
\quad , \quad  
\nabla_{\overline{P}} \mathcal{R} + 4 \nabla^{\underline{M}} \mathcal{R}_{\underline{M} \overline{P}} = 
0 \,.
\label{eq:BIsHZ}
\ee
The scalar curvature - which is the Lagrangian of double field theory - is given by $\mathcal{R} =
P^{MP} P^{NQ} \mathcal{R}_{MNPQ}$. 

\subsubsection{Generalised diffeomorphisms and Killing vectors}

As the connection is generalised torsion free, we can covariantise generalised diffeomorphisms
simply by replacing partial derivatives with covariant ones. Thus in particular
\be
\delta_\xi \gM^{MN} = - 4 \gM^{P(Q} \nabla_P \xi^{R)} \left( P_R^{M} \bar{P}_Q^{N} + \bar P_R^{M}
{P}_Q^{N} \right) \,,
\ee
\be
\delta_\xi d = - \frac{1}{2} \nabla_M \xi^M \,.
\ee
Thus a generalised Killing vector obeys
\be
\nabla_M \xi^M = 0 \quad , \quad
\gM^{P(Q} \nabla_P \xi^{R)} \left( P_R^{M} \bar{P}_Q^{N} + \bar P_R^{M}
{P}_Q^{N} \right) = 0 \,.
\ee

\subsection{The Weitzenb\"ock connection} 

\subsubsection{The connection}

This connection is given simply in terms of the generalised vielbein as
\be
\Omega_{MN}{}^P = E_\alpha{}^P \partial_M E^\alpha{}_N \,.
\ee

\subsubsection{Generalised Riemann tensor, torsion and Ricci identities}

The Weitzenb\"ock connection has vanishing generalised Riemann tensor (using section condition), but non-vanishing generalised torsion. It obeys the following generalised Ricci identities:
To simplify this we can use the Ricci-type identities
\be
\left( D_M \nabla_N - D_N \nabla_M \right) T  = -\tau_{MN}{}^U D_U T \,,
\ee
where $T$ is any generalised tensor, and
\be
\left(D_M D_N - D_N D_M \right) d  =  \frac{1}{2} \left( D_P - 2 D_P d
\right) \tau_{MN}{}^P \,.
\ee
In the paper \cite{Geissbuhler:2013uka}, this is actually viewed as a Bianchi identity.

\subsubsection{Generalised diffeomorphisms and Killing vectors}

As the connection has generalised torsion, one has 
\be
\delta_\xi V^M = \xi^N D_N V^M - V^N D_N \xi^M + \eta^{MN} \eta_{PQ} D_N \xi^P V^Q
+ \tau_{NP}{}^M V^N \xi^P \,,
\ee
so in particular 
\be
\delta_\xi \gM^{MN} = - 2 \gM^{P(M } D_P \xi^{N)} + 2 \eta^{P(M} \eta_{QR} H^{N) Q} D_P \xi^R
+ 2 \gM^{P(M} \tau_{PQ}{}^{N)} \xi^Q \,,
\ee
and
\be
\delta_\xi d = \xi^M D_M d - \frac{1}{2} D_M \xi^M \,.
\ee
Hence a generalised Killing vector obeys
\be
- 2 \gM^{P(M } D_P \xi^{N)} + 2 \eta^{P(M} \eta_{QR} H^{N) Q} D_P \xi^R
+ 2 \gM^{P(M} \tau_{PQ}{}^{N)} \xi^Q = 0 \,,
\ee
and
\be
D_M \xi^M =  2 \xi^M D_M d  \,.
\ee

\subsubsection{Flux content of the generalised torsion}
\label{appflux}

Following \cite{Geissbuhler:2013uka}, we can write down the most general vielbein $E^\alpha{}_M$ in components
\be
E^\mu{}_i =  e^\mu{}_i  + e^\mu{}_l \beta^{lm} B_{mi}\quad , \quad 
E_\mu{}_i = B_{il} e_\mu{}^l  \quad , \quad 
E^\mu{}^i = \beta^{il} e^\mu{}_l  \quad , \quad 
E_\mu{}^i = e_\mu{}^i \,.
\label{eq:genviel}
\ee
These are also the components of the inverse vielbein. We see that this involves both the $B$-field
and the bivector ($e_\mu{}^i$ is the spacetime vielbein). Ordinarily, one will fix the local $O(D) \times O(D)$ by setting $\beta^{ij} = 0$. This defines a particular choice of the fields in the physical frame. 

Using \eqref{eq:genviel} one work out the torsion components
 %
\be
\tau_{ijk} = 3 D_{[i} B_{jk]} + 3 D_{[i} \beta^{lm} B_{j|l|} B_{k]m} \,,
\ee
\be
\tau_{ij}{}^k = T_{ij}{}^k + 2 B_{l[i} D_{j]} \beta^{lk} + \widetilde D^k B_{ij} + B_{il}
B_{jm} \widetilde D^k \beta^{lm} \,,
\ee
\be
\tau_i{}^{jk} =  D_i \beta^{jk} + 2 \widetilde\Gamma^{[k}{}_i{}^{j]} + 2 B_{il}
\widetilde D^{[j} \beta^{k]l} \,,
\ee
\be
\tau^{ijk} =  3 \widetilde D^{[i} \beta^{jk]} \,.
\ee
Here tildes are used to denote the derivatives with respect to dual directions (at this point we have not explicitly imposed the section condition). In spacetime we have $\Gamma_{Mi}{}^j = e^j_\mu \partial_M e_i^\mu$ and $D$ denotes the covariant derivative
with respect to this. 
%
We see that $\tau_{ijk}$ contains the usual $H$-flux, $H_{ijk} = 3 \partial_{[i} B_{jk]}$, $\tau_{ij}{}^k$ the usual geometric flux, $T_{ij}{}^k$, $\tau_i{}^{jk}$ the $Q$-flux $Q_i{}^{jk} = \partial_i \beta^{jk}$ and $\tau^{ijk}$ the $R$-flux, $R^{ijk} = 3 \tilde\partial^{[i} \beta^{jk]}$.

\bibliographystyle{JHEP}

\begin{thebibliography}{10}

\bibitem{Hitchin:2004ut}
N.~Hitchin, {\it {Generalized Calabi-Yau manifolds}},  {\em Quart.J.Math.Oxford
  Ser.} {\bf 54} (2003) 281--308,
  [\href{http://arxiv.org/abs/math/0209099}{{\tt math/0209099}}].

\bibitem{Gualtieri:2003dx}
M.~Gualtieri, {\it {Generalized complex geometry}},
  \href{http://arxiv.org/abs/math/0401221}{{\tt math/0401221}}.

\bibitem{Coimbra:2011nw}
A.~Coimbra, C.~Strickland-Constable, and D.~Waldram, {\it {Supergravity as
  Generalised Geometry I: Type II Theories}},  {\em JHEP} {\bf 1111} (2011)
  091, [\href{http://arxiv.org/abs/1107.1733}{{\tt arXiv:1107.1733}}].

\bibitem{Coimbra:2012yy}
A.~Coimbra, C.~Strickland-Constable, and D.~Waldram, {\it {Generalised Geometry
  and type II Supergravity}},  {\em Fortsch.Phys.} {\bf 60} (2012) 982--986,
  [\href{http://arxiv.org/abs/1202.3170}{{\tt arXiv:1202.3170}}].

\bibitem{Duff:1989tf}
M.~Duff, {\it {Duality rotations in string theory}},  {\em Nucl.Phys.} {\bf
  B335} (1990) 610.

\bibitem{Tseytlin:1990nb}
A.~A. Tseytlin, {\it {Duality symmetric formulation of string world sheet
  dynamics}},  {\em Phys.Lett.} {\bf B242} (1990) 163--174.

\bibitem{Tseytlin:1990va}
A.~A. Tseytlin, {\it {Duality symmetric closed string theory and interacting
  chiral scalars}},  {\em Nucl.Phys.} {\bf B350} (1991) 395--440.

\bibitem{Siegel:1993xq}
W.~Siegel, {\it {Two vierbein formalism for string inspired axionic gravity}},
  {\em Phys.Rev.} {\bf D47} (1993) 5453--5459,
  [\href{http://arxiv.org/abs/hep-th/9302036}{{\tt hep-th/9302036}}].

\bibitem{Siegel:1993th}
W.~Siegel, {\it {Superspace duality in low-energy superstrings}},  {\em
  Phys.Rev.} {\bf D48} (1993) 2826--2837,
  [\href{http://arxiv.org/abs/hep-th/9305073}{{\tt hep-th/9305073}}].

\bibitem{Hull:2009mi}
C.~Hull and B.~Zwiebach, {\it {Double Field Theory}},  {\em JHEP} {\bf 0909}
  (2009) 099, [\href{http://arxiv.org/abs/0904.4664}{{\tt arXiv:0904.4664}}].

\bibitem{Hull:2009zb}
C.~Hull and B.~Zwiebach, {\it {The Gauge algebra of double field theory and
  Courant brackets}},  {\em JHEP} {\bf 0909} (2009) 090,
  [\href{http://arxiv.org/abs/0908.1792}{{\tt arXiv:0908.1792}}].

\bibitem{Hohm:2010jy}
O.~Hohm, C.~Hull, and B.~Zwiebach, {\it {Background independent action for
  double field theory}},  {\em JHEP} {\bf 1007} (2010) 016,
  [\href{http://arxiv.org/abs/1003.5027}{{\tt arXiv:1003.5027}}].

\bibitem{Hohm:2010pp}
O.~Hohm, C.~Hull, and B.~Zwiebach, {\it {Generalized metric formulation of
  double field theory}},  {\em JHEP} {\bf 1008} (2010) 008,
  [\href{http://arxiv.org/abs/1006.4823}{{\tt arXiv:1006.4823}}].

\bibitem{Hohm:2010xe}
O.~Hohm and S.~K. Kwak, {\it {Frame-like Geometry of Double Field Theory}},
  {\em J.Phys.} {\bf A44} (2011) 085404,
  [\href{http://arxiv.org/abs/1011.4101}{{\tt arXiv:1011.4101}}].

\bibitem{Hohm:2011zr}
O.~Hohm, S.~K. Kwak, and B.~Zwiebach, {\it {Unification of Type II Strings and
  T-duality}},  {\em Phys.Rev.Lett.} {\bf 107} (2011) 171603,
  [\href{http://arxiv.org/abs/1106.5452}{{\tt arXiv:1106.5452}}].

\bibitem{Hohm:2011dv}
O.~Hohm, S.~K. Kwak, and B.~Zwiebach, {\it {Double Field Theory of Type II
  Strings}},  {\em JHEP} {\bf 1109} (2011) 013,
  [\href{http://arxiv.org/abs/1107.0008}{{\tt arXiv:1107.0008}}].

\bibitem{Hohm:2011ex}
O.~Hohm and S.~K. Kwak, {\it {Double Field Theory Formulation of Heterotic
  Strings}},  {\em JHEP} {\bf 1106} (2011) 096,
  [\href{http://arxiv.org/abs/1103.2136}{{\tt arXiv:1103.2136}}].

\bibitem{Hohm:2011cp}
O.~Hohm and S.~K. Kwak, {\it {Massive Type II in Double Field Theory}},  {\em
  JHEP} {\bf 1111} (2011) 086, [\href{http://arxiv.org/abs/1108.4937}{{\tt
  arXiv:1108.4937}}].

\bibitem{Hohm:2011nu}
O.~Hohm and S.~K. Kwak, {\it {N=1 Supersymmetric Double Field Theory}},  {\em
  JHEP} {\bf 1203} (2012) 080, [\href{http://arxiv.org/abs/1111.7293}{{\tt
  arXiv:1111.7293}}].

\bibitem{Hohm:2013nja}
O.~Hohm and H.~Samtleben, {\it {Gauge theory of Kaluza-Klein and winding
  modes}},  {\em Phys.Rev.} {\bf D88} (2013) 085005,
  [\href{http://arxiv.org/abs/1307.0039}{{\tt arXiv:1307.0039}}].

\bibitem{Jeon:2011vx}
I.~Jeon, K.~Lee, and J.-H. Park, {\it {Incorporation of fermions into double
  field theory}},  {\em JHEP} {\bf 1111} (2011) 025,
  [\href{http://arxiv.org/abs/1109.2035}{{\tt arXiv:1109.2035}}].

\bibitem{Jeon:2011sq}
I.~Jeon, K.~Lee, and J.-H. Park, {\it {Supersymmetric Double Field Theory:
  Stringy Reformulation of Supergravity}},  {\em Phys.Rev.} {\bf D85} (2012)
  081501, [\href{http://arxiv.org/abs/1112.0069}{{\tt arXiv:1112.0069}}].

\bibitem{Jeon:2012kd}
I.~Jeon, K.~Lee, and J.-H. Park, {\it {Ramond-Ramond Cohomology and O(D,D)
  T-duality}},  {\em JHEP} {\bf 1209} (2012) 079,
  [\href{http://arxiv.org/abs/1206.3478}{{\tt arXiv:1206.3478}}].

\bibitem{Jeon:2012hp}
I.~Jeon, K.~Lee, J.-H. Park, and Y.~Suh, {\it {Stringy Unification of Type IIA
  and IIB Supergravities under N=2 D=10 Supersymmetric Double Field Theory}},
  {\em Phys.Lett.} {\bf B723} (2013) 245--250,
  [\href{http://arxiv.org/abs/1210.5078}{{\tt arXiv:1210.5078}}].

\bibitem{Jeon:2010rw}
I.~Jeon, K.~Lee, and J.-H. Park, {\it {Differential geometry with a projection:
  Application to double field theory}},  {\em JHEP} {\bf 1104} (2011) 014,
  [\href{http://arxiv.org/abs/1011.1324}{{\tt arXiv:1011.1324}}].

\bibitem{Jeon:2011cn}
I.~Jeon, K.~Lee, and J.-H. Park, {\it {Stringy differential geometry, beyond
  Riemann}},  {\em Phys.Rev.} {\bf D84} (2011) 044022,
  [\href{http://arxiv.org/abs/1105.6294}{{\tt arXiv:1105.6294}}].

\bibitem{Hohm:2011si}
O.~Hohm and B.~Zwiebach, {\it {On the Riemann Tensor in Double Field Theory}},
  {\em JHEP} {\bf 1205} (2012) 126, [\href{http://arxiv.org/abs/1112.5296}{{\tt
  arXiv:1112.5296}}].

\bibitem{Hohm:2012mf}
O.~Hohm and B.~Zwiebach, {\it {Towards an invariant geometry of double field
  theory}},  {\em J.Math.Phys.} {\bf 54} (2013) 032303,
  [\href{http://arxiv.org/abs/1212.1736}{{\tt arXiv:1212.1736}}].

\bibitem{Berman:2013uda}
D.~S. Berman, C.~D.~A. Blair, E.~Malek, and M.~J. Perry, {\it {The $O_{D,D}$
  geometry of string theory}},  {\em Int.J.Mod.Phys.} {\bf A29} (2014), no.~15
  1450080, [\href{http://arxiv.org/abs/1303.6727}{{\tt arXiv:1303.6727}}].

\bibitem{Grana:2012rr}
M.~Gra{\~n}a and D.~Marqu{\'e}s, {\it {Gauged Double Field Theory}},  {\em
  JHEP} {\bf 1204} (2012) 020, [\href{http://arxiv.org/abs/1201.2924}{{\tt
  arXiv:1201.2924}}].

\bibitem{Aldazabal:2011nj}
G.~Aldazabal, W.~Baron, D.~Marqu{\'e}s, and C.~N{\'u}{\~n}ez, {\it {The
  effective action of Double Field Theory}},  {\em JHEP} {\bf 1111} (2011) 052,
  [\href{http://arxiv.org/abs/1109.0290}{{\tt arXiv:1109.0290}}].

\bibitem{Berman:2013cli}
D.~S. Berman and K.~Lee, {\it {Supersymmetry for Gauged Double Field Theory and
  Generalised Scherk-Schwarz Reductions}},
  \href{http://arxiv.org/abs/1305.2747}{{\tt arXiv:1305.2747}}.

\bibitem{Berman:2011kg}
D.~S. Berman, E.~T. Musaev, and M.~J. Perry, {\it {Boundary Terms in
  Generalized Geometry and doubled field theory}},  {\em Phys.Lett.} {\bf B706}
  (2011) 228--231, [\href{http://arxiv.org/abs/1110.3097}{{\tt
  arXiv:1110.3097}}].

\bibitem{Geissbuhler:2011mx}
D.~Geissb{\"u}hler, {\it {Double Field Theory and N=4 Gauged Supergravity}},
  {\em JHEP} {\bf 1111} (2011) 116, [\href{http://arxiv.org/abs/1109.4280}{{\tt
  arXiv:1109.4280}}].

\bibitem{Geissbuhler:2013uka}
D.~Geissb{\"u}hler, D.~Marqu{\'e}s, C.~N{\'u}{\~n}ez, and V.~Penas, {\it
  {Exploring Double Field Theory}},  {\em JHEP} {\bf 1306} (2013) 101,
  [\href{http://arxiv.org/abs/1304.1472}{{\tt arXiv:1304.1472}}].

\bibitem{Hohm:2012gk}
O.~Hohm and B.~Zwiebach, {\it {Large Gauge Transformations in Double Field
  Theory}},  {\em JHEP} {\bf 1302} (2013) 075,
  [\href{http://arxiv.org/abs/1207.4198}{{\tt arXiv:1207.4198}}].

\bibitem{Park:2013mpa}
J.-H. Park, {\it {Comments on double field theory and diffeomorphisms}},  {\em
  JHEP} {\bf 1306} (2013) 098, [\href{http://arxiv.org/abs/1304.5946}{{\tt
  arXiv:1304.5946}}].

\bibitem{Berman:2014jba}
D.~S. Berman, M.~Cederwall, and M.~J. Perry, {\it {Global aspects of double
  geometry}},  {\em JHEP} {\bf 1409} (2014) 066,
  [\href{http://arxiv.org/abs/1401.1311}{{\tt arXiv:1401.1311}}].

\bibitem{Papadopoulos:2014mxa}
G.~Papadopoulos, {\it {Seeking the balance: Patching double and exceptional
  field theories}},  \href{http://arxiv.org/abs/1402.2586}{{\tt
  arXiv:1402.2586}}.

\bibitem{Cederwall:2014kxa}
M.~Cederwall, {\it {The geometry behind double geometry}},  {\em JHEP} {\bf
  1409} (2014) 070, [\href{http://arxiv.org/abs/1402.2513}{{\tt
  arXiv:1402.2513}}].

\bibitem{Cederwall:2014opa}
M.~Cederwall, {\it {T-duality and non-geometric solutions from double
  geometry}},  \href{http://arxiv.org/abs/1409.4463}{{\tt arXiv:1409.4463}}.

\bibitem{Andriot:2012an}
D.~Andriot, O.~Hohm, M.~Larfors, D.~L{\"u}st, and P.~Patalong, {\it
  {Non-Geometric Fluxes in Supergravity and Double Field Theory}},  {\em
  Fortsch.Phys.} {\bf 60} (2012) 1150--1186,
  [\href{http://arxiv.org/abs/1204.1979}{{\tt arXiv:1204.1979}}].

\bibitem{Berkeley:2014nza}
J.~Berkeley, D.~S. Berman, and F.~J. Rudolph, {\it {Strings and Branes are
  Waves}},  \href{http://arxiv.org/abs/1403.7198}{{\tt arXiv:1403.7198}}.

\bibitem{Berman:2014jsa}
D.~S. Berman and F.~J. Rudolph, {\it {Branes are Waves and Monopoles}},
  \href{http://arxiv.org/abs/1409.6314}{{\tt arXiv:1409.6314}}.

\bibitem{Blumenhagen:2014gva}
R.~Blumenhagen, F.~Hassler, and D.~Lust, {\it {Double Field Theory on Group
  Manifolds}},  {\em JHEP} {\bf 02} (2015) 001,
  [\href{http://arxiv.org/abs/1410.6374}{{\tt arXiv:1410.6374}}].

\bibitem{Blumenhagen:2015zma}
R.~Blumenhagen, P.~d. Bosque, F.~Hassler, and D.~Lust, {\it {Generalized Metric
  Formulation of Double Field Theory on Group Manifolds}},
  \href{http://arxiv.org/abs/1502.02428}{{\tt arXiv:1502.02428}}.

\bibitem{Aldazabal:2013sca}
G.~Aldazabal, D.~Marqu{\'e}s, and C.~N{\'u}{\~n}ez, {\it {Double Field Theory:
  A Pedagogical Review}},  {\em Class.Quant.Grav.} {\bf 30} (2013) 163001,
  [\href{http://arxiv.org/abs/1305.1907}{{\tt arXiv:1305.1907}}].

\bibitem{Berman:2013eva}
D.~S. Berman and D.~C. Thompson, {\it {Duality Symmetric String and M-Theory}},
   \href{http://arxiv.org/abs/1306.2643}{{\tt arXiv:1306.2643}}.

\bibitem{Hohm:2013bwa}
O.~Hohm, D.~L{\"u}st, and B.~Zwiebach, {\it {The Spacetime of Double Field
  Theory: Review, Remarks, and Outlook}},  {\em Fortsch.Phys.} {\bf 61} (2013)
  926--966, [\href{http://arxiv.org/abs/1309.2977}{{\tt arXiv:1309.2977}}].

\bibitem{Park:2015bza}
J.-H. Park, S.-J. Rey, W.~Rim, and Y.~Sakatani, {\it {O(D,D) Covariant Noether
  Currents and Global Charges in Double Field Theory}},
  \href{http://arxiv.org/abs/1507.07545}{{\tt arXiv:1507.07545}}.

\bibitem{Ortin:2004ms}
T.~Ortin, {\em {Gravity and strings}}.
\newblock Cambridge Univ. Press, 2004.

\bibitem{Kwak:2010ew}
S.~K. Kwak, {\it {Invariances and Equations of Motion in Double Field Theory}},
   {\em JHEP} {\bf 10} (2010) 047, [\href{http://arxiv.org/abs/1008.2746}{{\tt
  arXiv:1008.2746}}].

\bibitem{Naseer:2015fba}
U.~Naseer, {\it {Canonical formulation and conserved charges of double field
  theory}},  \href{http://arxiv.org/abs/1508.00844}{{\tt arXiv:1508.00844}}.

\bibitem{Andriot:2011uh}
D.~Andriot, M.~Larfors, D.~L{\"u}st, and P.~Patalong, {\it {A ten-dimensional
  action for non-geometric fluxes}},  {\em JHEP} {\bf 1109} (2011) 134,
  [\href{http://arxiv.org/abs/1106.4015}{{\tt arXiv:1106.4015}}].

\bibitem{Andriot:2012wx}
D.~Andriot, O.~Hohm, M.~Larfors, D.~L{\"u}st, and P.~Patalong, {\it {A
  geometric action for non-geometric fluxes}},  {\em Phys.Rev.Lett.} {\bf 108}
  (2012) 261602, [\href{http://arxiv.org/abs/1202.3060}{{\tt
  arXiv:1202.3060}}].

\bibitem{Andriot:2013xca}
D.~Andriot and A.~Betz, {\it {$\beta$-supergravity: a ten-dimensional theory
  with non-geometric fluxes, and its geometric framework}},  {\em JHEP} {\bf
  1312} (2013) 083, [\href{http://arxiv.org/abs/1306.4381}{{\tt
  arXiv:1306.4381}}].

\bibitem{deBoer:2010ud}
J.~de~Boer and M.~Shigemori, {\it {Exotic branes and non-geometric
  backgrounds}},  {\em Phys.Rev.Lett.} {\bf 104} (2010) 251603,
  [\href{http://arxiv.org/abs/1004.2521}{{\tt arXiv:1004.2521}}].

\bibitem{deBoer:2012ma}
J.~de~Boer and M.~Shigemori, {\it {Exotic Branes in String Theory}},  {\em
  Phys.Rept.} {\bf 532} (2013) 65--118,
  [\href{http://arxiv.org/abs/1209.6056}{{\tt arXiv:1209.6056}}].

\bibitem{Bergshoeff:2011se}
E.~A. Bergshoeff, T.~Ort{\'i}n, and F.~Riccioni, {\it {Defect Branes}},  {\em
  Nucl.Phys.} {\bf B856} (2012) 210--227,
  [\href{http://arxiv.org/abs/1109.4484}{{\tt arXiv:1109.4484}}].

\bibitem{Sakatani:2014hba}
Y.~Sakatani, {\it {Exotic branes and non-geometric fluxes}},  {\em JHEP} {\bf
  03} (2015) 135, [\href{http://arxiv.org/abs/1412.8769}{{\tt
  arXiv:1412.8769}}].

\bibitem{Berman:2014hna}
D.~S. Berman and F.~J. Rudolph, {\it {Strings, Branes and the Self-dual
  Solutions of Exceptional Field Theory}},  {\em JHEP} {\bf 05} (2015) 130,
  [\href{http://arxiv.org/abs/1412.2768}{{\tt arXiv:1412.2768}}].

\bibitem{Malek:2013sp}
E.~Malek, {\it {Timelike U-dualities in Generalised Geometry}},  {\em JHEP}
  {\bf 1311} (2013) 185, [\href{http://arxiv.org/abs/1301.0543}{{\tt
  arXiv:1301.0543}}].

\bibitem{Ellwood:2006ya}
I.~T. Ellwood, {\it {NS-NS fluxes in Hitchin's generalized geometry}},  {\em
  JHEP} {\bf 0712} (2007) 084, [\href{http://arxiv.org/abs/hep-th/0612100}{{\tt
  hep-th/0612100}}].

\bibitem{Grana:2008yw}
M.~Gra{\~n}a, R.~Minasian, M.~Petrini, and D.~Waldram, {\it {T-duality,
  Generalized Geometry and Non-Geometric Backgrounds}},  {\em JHEP} {\bf 0904}
  (2009) 075, [\href{http://arxiv.org/abs/0807.4527}{{\tt arXiv:0807.4527}}].

\bibitem{Bergshoeff:2010xc}
E.~A. Bergshoeff and F.~Riccioni, {\it {D-Brane Wess-Zumino Terms and
  U-Duality}},  {\em JHEP} {\bf 11} (2010) 139,
  [\href{http://arxiv.org/abs/1009.4657}{{\tt arXiv:1009.4657}}].

\bibitem{Bergshoeff:2011zk}
E.~A. Bergshoeff and F.~Riccioni, {\it {String Solitons and T-duality}},  {\em
  JHEP} {\bf 1105} (2011) 131, [\href{http://arxiv.org/abs/1102.0934}{{\tt
  arXiv:1102.0934}}].

\bibitem{Bergshoeff:2011ee}
E.~A. Bergshoeff and F.~Riccioni, {\it {Branes and wrapping rules}},  {\em
  Phys.Lett.} {\bf B704} (2011) 367--372,
  [\href{http://arxiv.org/abs/1108.5067}{{\tt arXiv:1108.5067}}].

\bibitem{Bergshoeff:2012ex}
E.~A. Bergshoeff, A.~Marrani, and F.~Riccioni, {\it {Brane orbits}},  {\em
  Nucl. Phys.} {\bf B861} (2012) 104--132,
  [\href{http://arxiv.org/abs/1201.5819}{{\tt arXiv:1201.5819}}].

\bibitem{Berman:2010is}
D.~S. Berman and M.~J. Perry, {\it {Generalized Geometry and M theory}},  {\em
  JHEP} {\bf 1106} (2011) 074, [\href{http://arxiv.org/abs/1008.1763}{{\tt
  arXiv:1008.1763}}].

\bibitem{Aldazabal:2013mya}
G.~Aldazabal, M.~Gra{\~n}a, D.~Marqu{\'e}s, and J.~Rosabal, {\it {Extended
  geometry and gauged maximal supergravity}},  {\em JHEP} {\bf 1306} (2013)
  046, [\href{http://arxiv.org/abs/1302.5419}{{\tt arXiv:1302.5419}}].

\bibitem{Hohm:2013vpa}
O.~Hohm and H.~Samtleben, {\it {Exceptional Field Theory I: $E_{6(6)}$
  covariant Form of M-Theory and Type IIB}},  {\em Phys.Rev.} {\bf D89} (2014)
  066016, [\href{http://arxiv.org/abs/1312.0614}{{\tt arXiv:1312.0614}}].

\bibitem{Hohm:2013uia}
O.~Hohm and H.~Samtleben, {\it {Exceptional Field Theory II: E$_{7(7)}$}},
  {\em Phys.Rev.} {\bf D89} (2014) 066017,
  [\href{http://arxiv.org/abs/1312.4542}{{\tt arXiv:1312.4542}}].

\bibitem{Hohm:2014fxa}
O.~Hohm and H.~Samtleben, {\it {Exceptional Field Theory III: E$_{8(8)}$}},
  {\em Phys.Rev.} {\bf D90} (2014) 066002,
  [\href{http://arxiv.org/abs/1406.3348}{{\tt arXiv:1406.3348}}].

\bibitem{Godazgar:2014nqa}
H.~Godazgar, M.~Godazgar, O.~Hohm, H.~Nicolai, and H.~Samtleben, {\it
  {Supersymmetric E$_{7(7)}$ Exceptional Field Theory}},
  \href{http://arxiv.org/abs/1406.3235}{{\tt arXiv:1406.3235}}.

\bibitem{Blair:2014zba}
C.~D.~A. Blair and E.~Malek, {\it {Geometry and fluxes of SL(5) exceptional
  field theory}},  {\em JHEP} {\bf 1503} (2015) 144,
  [\href{http://arxiv.org/abs/1412.0635}{{\tt arXiv:1412.0635}}].

\bibitem{Page:1984qv}
D.~N. Page, {\it {Classical Stability of Round and Squashed Seven Spheres in
  Eleven-dimensional Supergravity}},  {\em Phys. Rev.} {\bf D28} (1983) 2976.

\bibitem{Marolf:2000cb}
D.~Marolf, {\it {Chern-Simons terms and the three notions of charge}},  in {\em
  {Quantization, gauge theory, and strings. Proceedings, International
  Conference dedicated to the memory of Professor Efim Fradkin, Moscow, Russia,
  June 5-10, 2000. Vol. 1+2}}, pp.~312--320, 2000.
\newblock \href{http://arxiv.org/abs/hep-th/0006117}{{\tt hep-th/0006117}}.

\bibitem{Kikuchi:2012za}
T.~Kikuchi, T.~Okada, and Y.~Sakatani, {\it {Rotating string in doubled
  geometry with generalized isometries}},  {\em Phys.Rev.} {\bf D86} (2012)
  046001, [\href{http://arxiv.org/abs/1205.5549}{{\tt arXiv:1205.5549}}].

\bibitem{Okada:2014wma}
T.~Okada and Y.~Sakatani, {\it {Defect branes as Alice strings}},  {\em JHEP}
  {\bf 03} (2015) 131, [\href{http://arxiv.org/abs/1411.1043}{{\tt
  arXiv:1411.1043}}].

\end{thebibliography}
\providecommand{\href}[2]{#2}\begingroup\raggedright\endgroup

\end{document}